\documentclass[11pt]{article}

\usepackage[preprint]{acl}

\usepackage{times}
\usepackage{latexsym}

\usepackage[T1]{fontenc}

\usepackage[utf8]{inputenc}

\usepackage{microtype}

\usepackage{inconsolata}

\usepackage{graphicx}
\usepackage{multirow}
\usepackage{xcolor}
\usepackage{amssymb}
\usepackage{booktabs}
\usepackage{makecell}
\usepackage{natbib}
\usepackage{stfloats}
\usepackage{float}
\usepackage{caption}
\usepackage[breakable]{tcolorbox}
\newtcolorbox{examplebox}{
  colback=gray!8, colframe=gray!80, boxrule=1pt,
  arc=2pt, left=4pt, right=4pt, top=4pt, bottom=4pt,
  breakable
}
\usepackage{cuted}
\title{EduIllustrate: Towards Scalable Automated Generation Of Multimodal Educational Content}

\author{
  \textbf{Shuzhen Bi\textsuperscript{1,2}},
  \textbf{Mingzi Zhang\textsuperscript{3}},
  \textbf{Zhuoxuan Li\textsuperscript{3}},
  \textbf{Xiaolong Wang\textsuperscript{3}},
  \textbf{Keqian Li\textsuperscript{3}\thanks{Corresponding author.}},
  \textbf{Aimin Zhou\textsuperscript{1,3}}
\\
\\
  \textsuperscript{1}Shanghai Innovation Institute,
  \textsuperscript{2}University of Science and Technology of China,
  \textsuperscript{3}East China Normal University
\\
  \small{
    \texttt{sa22916003@mail.ustc.edu.cn},
    \texttt{\{51284102005, lizhuoxuan, wmumu\}@stu.ecnu.edu.cn},
    \texttt{kqli@mail.ecnu.edu.cn},
    \texttt{amzhou@cs.ecnu.cn}
  }
}

\begin{document}
\maketitle

\begin{figure*}[t]
\centering
\includegraphics[width=\linewidth]{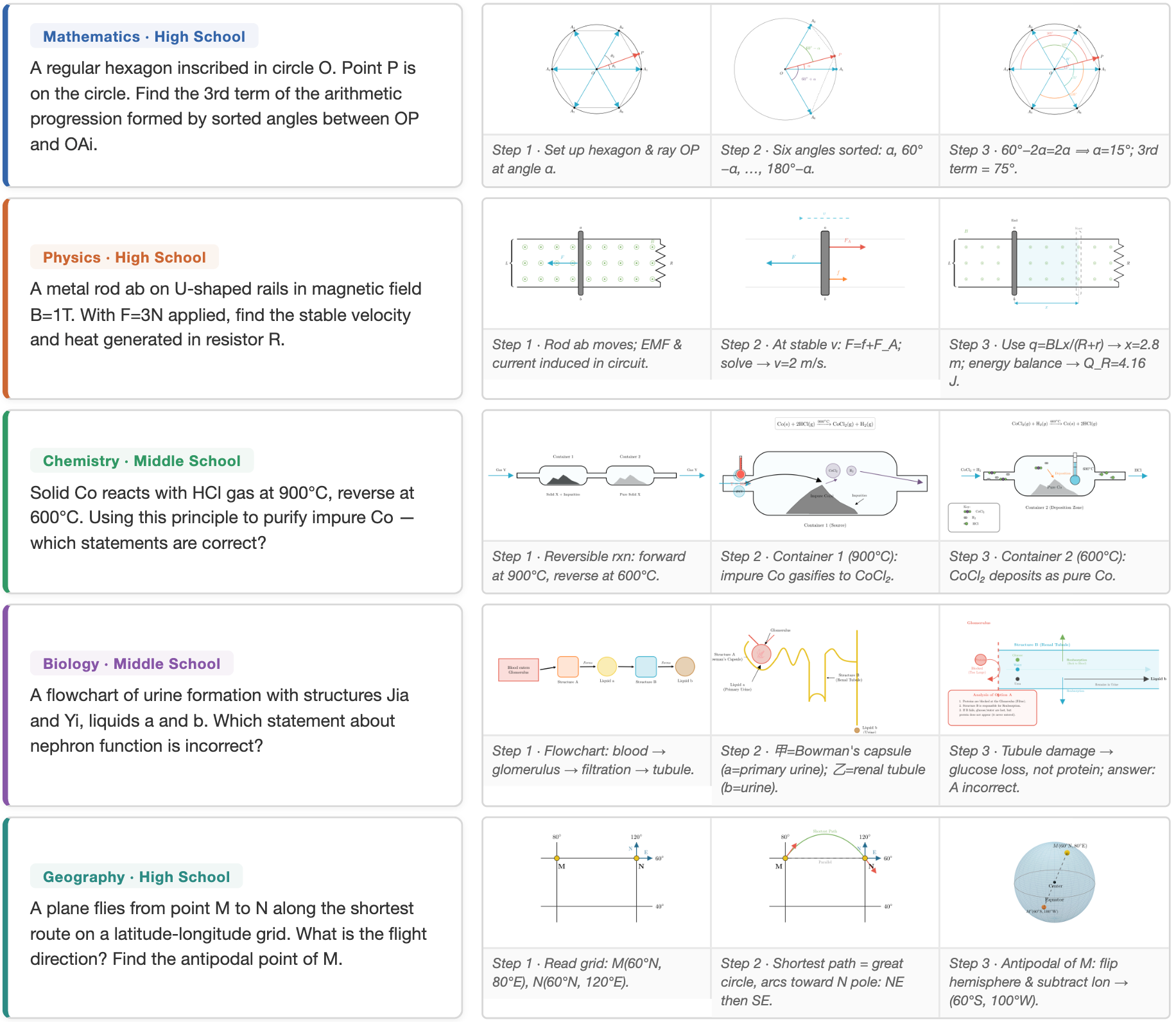}
\caption{EduIllustrate generates geometrically accurate visuals for diverse STEM problems. Taking textual K-12 problems from various subjects as input (left), our system produces a progressive sequence of diagrams (right). These diagrams are interleaved with textual reasoning to form multimodal explanations.}
\label{fig:teaser}
\end{figure*}

\begin{strip}
\centering
Website: \url{https://ecnu-innospark.github.io/EduIllustrate/}
Data:\url{
https://huggingface.co/datasets/keqianli/EduIllustrate}
\end{strip}

\begin{abstract}
Large language models are increasingly used as educational assistants, yet evaluation of their educational capabilities remains concentrated on question-answering and tutoring tasks. A critical gap exists for \textit{multimedia instructional content generation}---the ability to produce coherent, diagram-rich explanations that combine geometrically accurate visuals with step-by-step reasoning. We present \textbf{EduIllustrate}, a benchmark for evaluating LLMs on interleaved text-diagram explanation generation for K-12 STEM problems. The benchmark comprises \textbf{230 problems} spanning five subjects and three grade levels, a standardized generation protocol with sequential anchoring to enforce cross-diagram visual consistency, and an \textbf{8-dimension evaluation rubric} grounded in multimedia learning theory covering both text and visual quality. Evaluation of ten LLMs reveals a wide performance spread: Gemini 3.0 Pro Preview leads at 87.8\%, while Kimi-K2.5 achieves the best cost-efficiency (80.8\% at \$0.12/problem). Workflow ablation confirms sequential anchoring improves Visual Consistency by 13\% at 94\% lower cost. Human evaluation with 20 expert raters validates LLM-as-judge reliability for objective dimensions ($\rho \geq 0.83$) while revealing limitations on subjective visual assessment.
\end{abstract}

\section{Introduction}

\begin{figure*}[t]
  \centering
  \includegraphics[width=\textwidth]{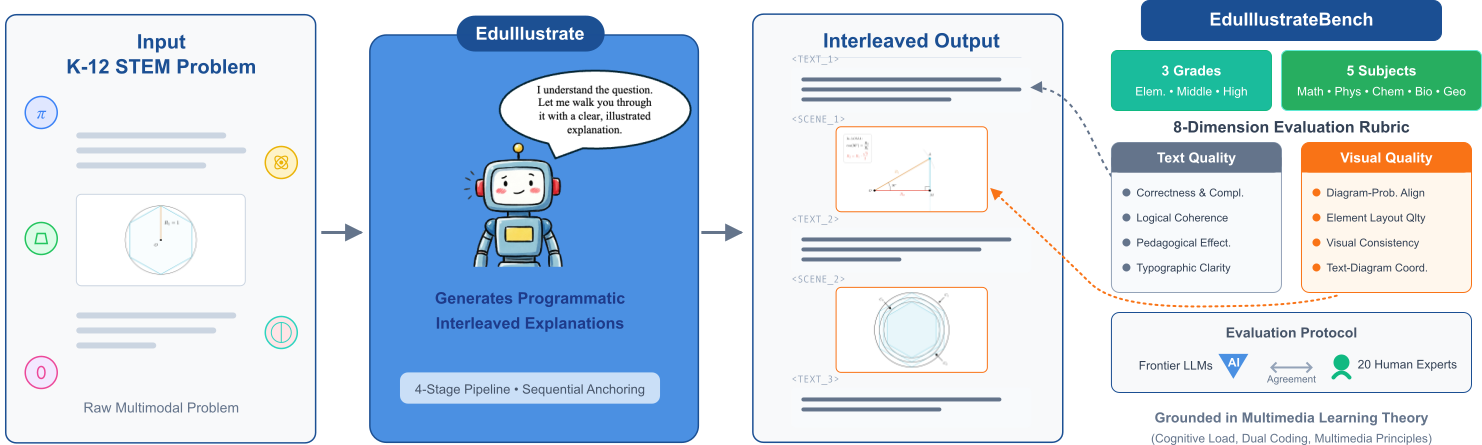}
  \caption{Overview of EduIllustrate. Given a K-12 STEM problem, the generation protocol produces a structured explanation document through four stages. The resulting output is evaluated using an 8-dimension rubric validated against human expert raters.}
  \label{fig:framework}
\end{figure*}

With the rapid development of large language models (LLMs), education has emerged as one of the most important and widely adopted application domains. LLMs are already used by millions of learners as on-demand educational assistants~\citep{kamalov2025agentic}, and major providers are actively building tutoring-oriented systems~\citep{wang2025genmentor,liu2025cogent}. However, prior evaluation of LLMs' educational capabilities has focused mainly on traditional question-answering and tutoring settings, leaving an important class of real educational tasks underexplored: the generation of \textit{multimedia instructional content}.

Decades of research in Cognitive Load Theory~\citep{sweller1988cognitive}, Dual Coding Theory~\citep{paivio1971imagery}, and the Multimedia Learning Principle~\citep{mayer2001multimedia} have established that coordinating visual diagrams with textual reasoning substantially reduces cognitive burden and deepens conceptual understanding. Yet producing such materials at scale remains a formidable challenge---lesson planning and material preparation rank among the most time-consuming aspects of teachers' professional work~\citep{philipp2013teachers,thompson2024teachers}, and many educators lack the specialized skills required to create geometrically accurate diagrams~\citep{turkozudincer2025literacy,yao2026instructional}. If LLMs could reliably generate illustrated explanations, the impact on K-12 education would be substantial. But we currently lack the benchmarks to measure whether they can.

Existing work falls short in several ways. On the education side, content generation systems bifurcate into text-only approaches~\citep{liu2025cogent,wang2025genmentor} and video-based systems targeting university-level theorems~\citep{ku2025theoremexplain,chen2025code2video}. On the multimodal generation side, systems such as ANOLE~\citep{chern2024anole} and Orthus~\citep{kou2025orthus} target general-purpose domains and generate photorealistic images, which cannot satisfy the geometric precision required for educational diagrams. Evaluation benchmarks like OpenLEAF~\citep{an2024openleaf} and MMIE~\citep{xia2025mmie} assess interleaved generation but not in educational contexts, while DiagramIR~\citep{kumar2025diagramir} evaluates mathematical diagrams in isolation and EduVisBench~\citep{ji2025eduvis} targets visual reasoning rather than generation quality. No existing benchmark jointly assesses both textual and visual quality of K-12 multimodal educational content.

To address this gap, we present \textbf{EduIllustrate} (Figure~\ref{fig:framework}), a benchmark designed to evaluate LLMs on interleaved text-diagram explanation generation for K-12 STEM problems---a setting that more closely reflects real-world educational applications. The benchmark comprises three components: (1) a curated problem set of 230 problems spanning five subjects and three grade levels; (2) a standardized generation protocol with sequential anchoring to ensure cross-diagram visual consistency; and (3) an \textbf{8-dimension evaluation rubric} grounded in multimedia learning theory that jointly covers textual and visual quality.

Our main contributions are threefold: (1) \textbf{EduIllustrateBench}, comprising 230 curated problems across five subjects and three grade levels, with an 8-dimension evaluation rubric addressing the gap in multi-subject, multi-grade multimodal educational content evaluation; (2) a \textbf{standardized generation protocol} with sequential anchoring to ensure cross-diagram visual consistency, serving as both the generation method and an ablation baseline; and (3) a \textbf{comprehensive empirical study} of ten LLMs spanning proprietary and open-weight models, with ablation studies and human evaluation validating LLM-as-judge reliability.

\section{Related Work}

\subsection{LLMs in Education}

LLM applications in education span intelligent tutoring, automated assessment, and adaptive content generation. \citet{kamalov2025agentic} reviewed agentic workflows in education, highlighting advancements in automated tutoring while noting constraints in adaptability that multi-agent frameworks address. \citet{wang2025genmentor} introduced GenMentor, an LLM-powered multi-agent framework for goal-oriented learning, demonstrating effective skill gap identification and personalized learning path scheduling.

For K-12-specific applications, \citet{liu2025cogent} proposed COGENT, a curriculum-oriented framework generating grade-appropriate content aligned with curriculum standards. Despite these advances, existing work predominantly focuses on text-only explanations or assessment, leaving multimodal content generation for K-12 STEM underexplored.

\subsection{Programmatic Educational Content Generation}

Recent work has explored generating educational content through executable code. \citet{ku2025theoremexplain} introduce an agentic approach for generating long-form theorem explanation videos using Manim animations, targeting university-level mathematics and physics. \citet{chen2025code2video} propose a code-centric agent framework for generating professional educational videos via executable Python code. Both systems target \textbf{video-based explanations} for advanced topics. Our work differs by focusing on \textbf{static diagram generation} for K-12 problem-solving explanations, where textbook-style illustrations enable self-paced learning across five subjects with subject-specific diagram conventions.

\subsection{Programmatic Diagram Generation}

Programmatic diagram generation leverages tools like TikZ, Manim, and Matplotlib. \citet{kumar2025diagramir} introduced DiagramIR, an automatic evaluation pipeline for educational math diagrams using intermediate representations of LaTeX TikZ code, demonstrating higher agreement with human raters than LLM-as-judge baselines. \citet{cui2025draw} proposed Draw with Thought, a training-free framework guiding multimodal LLMs to reconstruct scientific diagrams into editable mxGraph XML code through Chain-of-Thought reasoning.

\subsection{LLM-as-Judge Evaluation}

LLMs as evaluators provide practical alternatives to costly human evaluation. \citet{lee2025lemaj} revealed in the legal domain that reference-free evaluation protocols correlate better with human expert judgments. \citet{park2025agacci} demonstrated AGACCI framework improves accuracy through distributing specialized evaluation roles across collaborative agents.

\section{EduIllustrate Benchmark}

\subsection{Task Formulation}

Given a K-12 STEM problem (text and an optional diagram), a model must produce an interleaved explanation: a sequence of textual reasoning steps alternating with programmatically rendered diagrams. This task jointly demands (i) mathematically correct and pedagogically coherent text, (ii) geometrically accurate diagrams faithful to the problem setup, and (iii) visual consistency across multiple diagrams within the same explanation. Failure in any single aspect undermines the explanation's educational value, making this a challenging multimodal generation task.

\subsection{Problem Set}

\begin{figure*}[t]
  \centering
  \includegraphics[width=\textwidth]{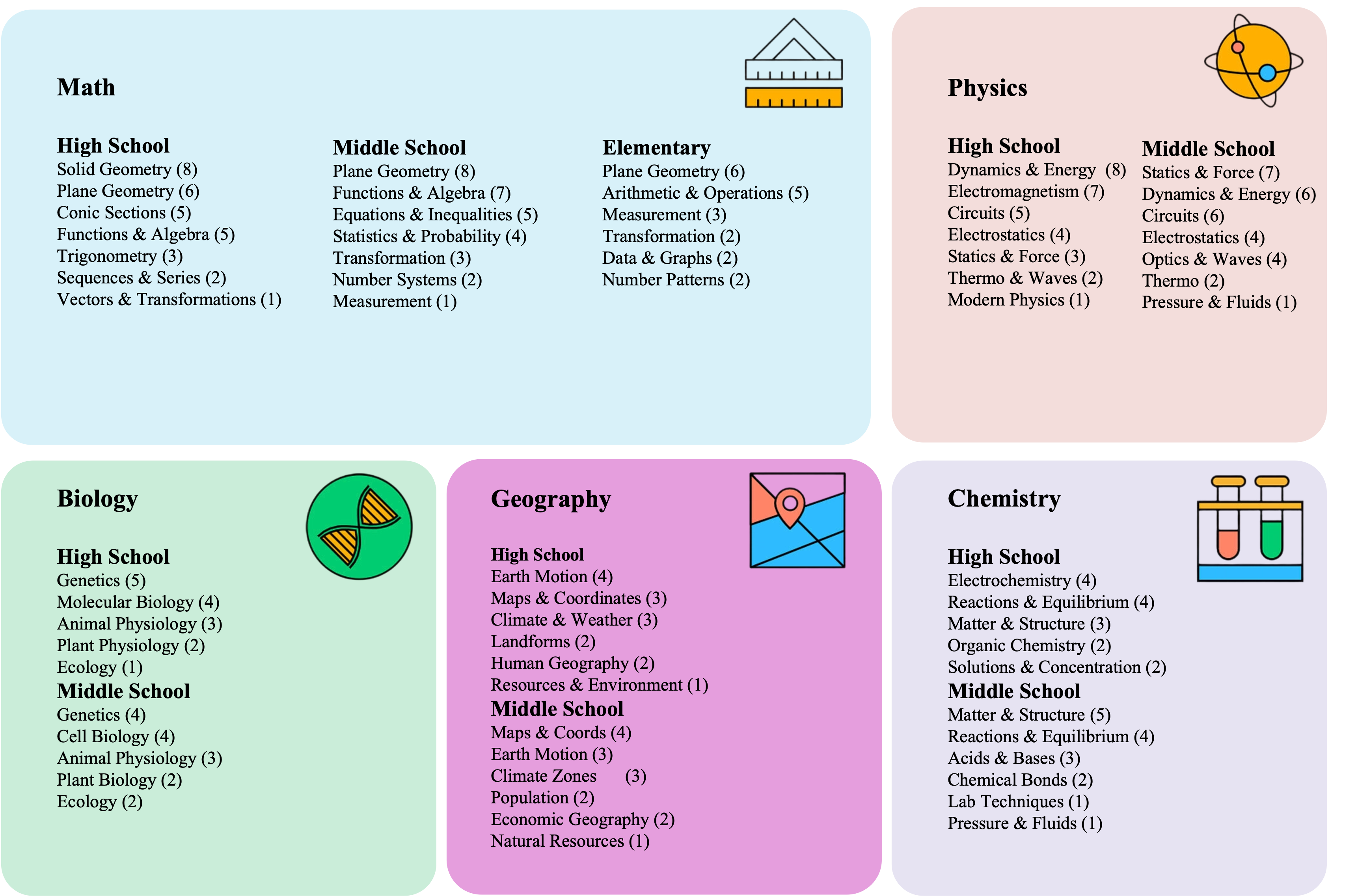}
  \caption{Topic categories by subject. Values in parentheses indicate the number of unique topics mapped in each category.}
  \label{fig:knowledge-categories}
\end{figure*}

We curate 230 problems from K12-Vista~\citep{zhang2025k12vista}, spanning three grade levels (elementary, middle, high school) and five STEM subjects (Table~\ref{tab:dataset}, Figure~\ref{fig:knowledge-categories}). Problems are selected for diagram appropriateness, solution clarity, and topic diversity; full curation details are provided in Appendix~\ref{sec:benchmark-details}.

\setlength{\tabcolsep}{4pt}
\begin{table}[t]
\centering
\small
\begin{tabular}{lcccccc}
\hline
\textbf{Grade} & \textbf{Math} & \textbf{Phys} & \textbf{Chem} & \textbf{Bio} & \textbf{Geo} & \textbf{Total} \\
\hline
Elementary & 20 & --- & --- & --- & --- & 20 \\
Middle     & 30 & 30  & 15  & 15  & 15  & 105 \\
High       & 30 & 30  & 15  & 15  & 15  & 105 \\
\hline
\textbf{Total} & \textbf{80} & \textbf{60} & \textbf{30} & \textbf{30} & \textbf{30} & \textbf{230} \\
\hline
\end{tabular}
\caption{Benchmark distribution across grade levels and subjects.}
\label{tab:dataset}
\end{table}

\subsection{Generation Protocol}

\begin{figure*}[t]
  \centering
  \includegraphics[width=\textwidth]{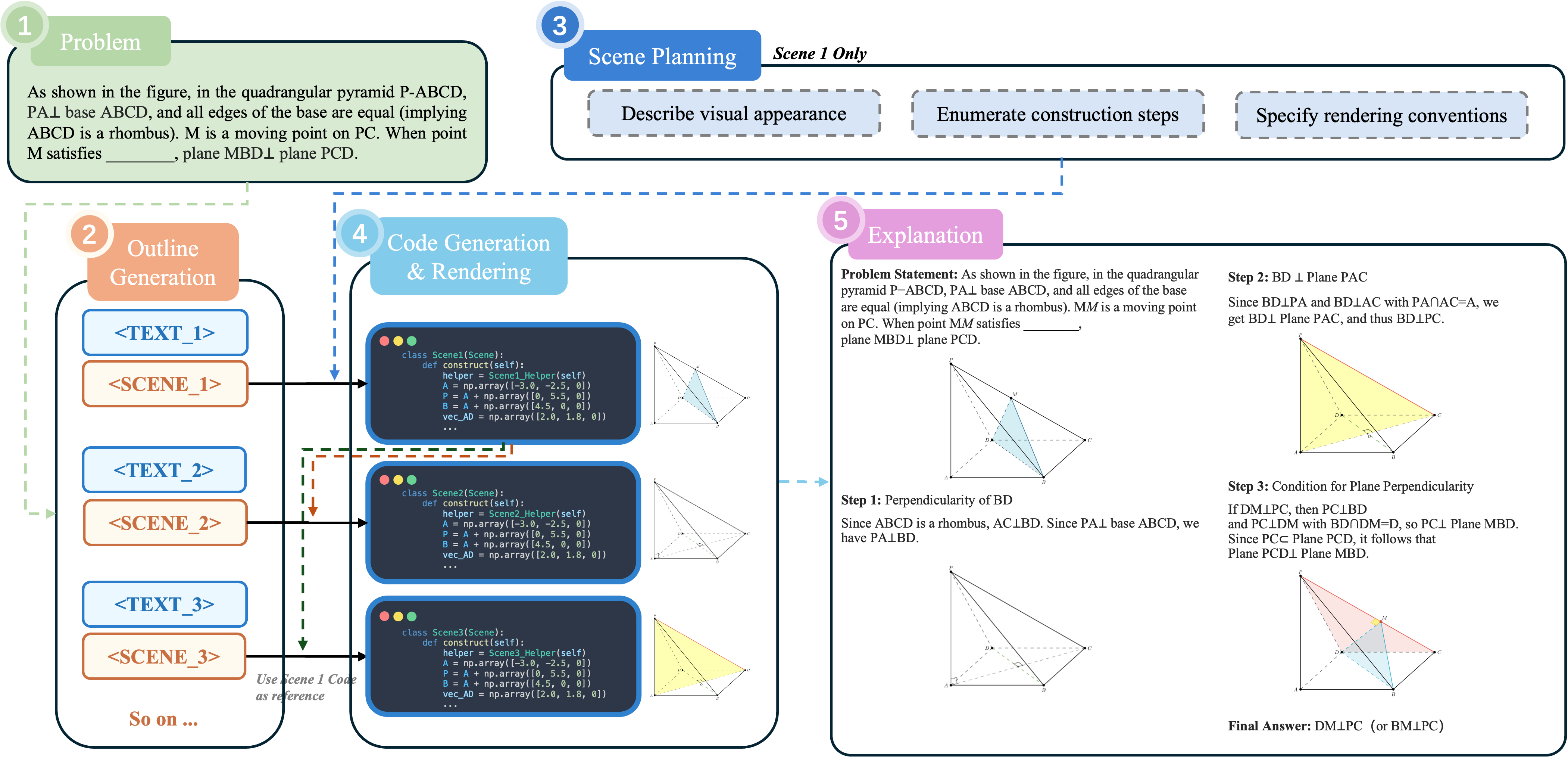}
  \caption{Generation protocol overview. Scene 1 undergoes full sequential processing (outline $\to$ implementation planning $\to$ code generation); Scenes 2--$N$ skip implementation planning and generate in parallel, each conditioned on Scene 1's code as a visual anchor.}
  \label{fig:pipeline}
\end{figure*}

To ensure fair comparison, all models generate explanations through a standardized four-stage protocol (Figure~\ref{fig:pipeline}). The protocol is designed to decouple the generation task into manageable subtasks while enforcing cross-diagram visual consistency.

\textbf{Stage 1: Structured Outline.} The model transforms a problem into an XML-based outline with alternating \texttt{<TEXT\_k>} (pedagogical explanation) and \texttt{<SCENE\_k>} (visual specification) blocks. Scene blocks are generated only when diagrams genuinely aid understanding. This structured format enables modular processing and error recovery.

\textbf{Stage 2: Implementation Planning (Scene 1 only).} Planning every scene independently would cause each to invent its own visual conventions, making consistency impossible to guarantee. We therefore restrict planning to Scene 1 only, converting its specification into a detailed implementation plan: intended appearance, spatial constraints, and discipline-specific rendering conventions (e.g., right-angle markers in geometry, vector arrows in physics). The resulting conventions---color scheme, labeling style, line weights---are inherited by all subsequent scenes.

\textbf{Stage 3: Code Generation and Rendering.} Scene 1 is rendered first from its plan; its complete Manim code then serves as context for Scenes 2--$N$, which generate in parallel. This enforces visual consistency without global optimization.

\textbf{Stage 4: Document Assembly.} Textual blocks and rendered images are assembled into a Markdown document in alternating order, requiring no LLM involvement.

\subsection{Evaluation Framework}

We propose an \textbf{8-dimension rubric} grounded in multimedia learning theory~\citep{sweller1988cognitive,paivio1971imagery,mayer2001multimedia}. Each dimension is scored on a 0--5 Likert scale and reported as a percentage (0--100\%).

\textit{Text quality} (evaluated on extracted \texttt{<TEXT\_k>} blocks): Correctness \& Completeness---whether the explanation reaches the correct answer through valid reasoning with all intermediate steps; Logical Coherence---whether reasoning flows naturally from premises to conclusions; Pedagogical Effectiveness---whether explanations use grade-appropriate language and effective instructional strategies; Typographic Clarity---proper mathematical notation, consistent formatting, and absence of artifacts.

\textit{Visual quality} (evaluated on rendered diagrams): Diagram--Problem Alignment---whether diagrams faithfully represent the problem's geometric or physical setup; Element Layout Quality---whether elements avoid overlap, maintain readable spacing, and follow discipline conventions; Visual Consistency---whether multiple diagrams maintain coherent color schemes, labeling, and style; Text--Diagram Coordination---how well textual references integrate with the corresponding diagrams.

\textbf{Automated Evaluation Protocol.} We employ Gemini 3.0 Pro Preview (temperature=0) as the judge model, selected for its strong multimodal reasoning and 2M-token context window. Evaluation input varies by dimension type: text-only dimensions receive the problem statement, extracted text, and rubric criteria (plus the gold solution for Correctness \& Completeness); multimodal dimensions additionally receive the relevant rendered diagrams alongside surrounding text. For Element Layout Quality, each diagram is scored independently and aggregated via geometric mean $ELQ = \left(\prod_{i=1}^N s_i\right)^{1/N}$, ensuring a single poor diagram substantially penalizes the overall score. For Visual Consistency, Scene 1 serves as the visual anchor; each subsequent diagram is compared against it, with $VC = \left(\prod_{i=2}^N s_{6,i}\right)^{1/(N-1)}$.

\textbf{Human Evaluation Protocol.} To validate LLM-as-judge reliability, we conduct human evaluation on 30 stratified random sampled explanations, scored by 20 expert raters across 7 dimensions (Correctness \& Completeness excluded) on a 3-level scale \{0, 0.5, 1\}, producing 4,200 judgments. To enable direct comparison with the automated scores (reported as percentages), LLM scores are normalized to [0, 1] by dividing by 5 before computing Spearman's $\rho$. We compute Krippendorff's $\alpha$ for inter-rater agreement and Spearman's $\rho$ for human-AI correlation. Full annotation procedures are described in Appendix~\ref{sec:annotation}.

\section{Experimental Results}

We evaluate ten LLMs on the EduIllustrate benchmark, analyzing performance across dimensions, subjects, and grade levels. We validate our sequential workflow design through ablation studies and assess LLM-as-judge reliability via human evaluation with 20 expert raters.

\subsection{Experimental Setup}

\textbf{Models Evaluated.} We evaluate ten LLMs spanning proprietary and open-weight categories: (1) Gemini 3.0 Pro Preview (Google DeepMind), (2) GPT-5 (OpenAI), (3) Claude Sonnet 4.5 (Anthropic), (4) Kimi-K2.5 (Moonshot AI), (5) Qwen3.5-397B, (6) Qwen3.5-122B, (7) Qwen 3.5-35B (Alibaba Cloud), (8) Mistral-Large-3, (9) Mistral-Small-4, and (10) Ministral-3-14B (Mistral AI). All models use identical prompts with temperature=0.7 on all 230 benchmark problems.

\subsection{Main Results}

\begin{table*}[t]
\centering
\renewcommand\arraystretch{1.3}
\setlength{\tabcolsep}{4pt}
\resizebox{\linewidth}{!}{\begin{tabular}{lcccc|cccc|cc}
\toprule
 & \multicolumn{4}{c|}{\textbf{Text Quality}} & \multicolumn{4}{c|}{\textbf{Visual Quality}} & & \\
\cmidrule(lr){2-5}\cmidrule(lr){6-9}
\textbf{Model} &
\makecell{\small\textbf{Correctness \&}\\\small\textbf{Completeness}} &
\makecell{\small\textbf{Logical}\\\small\textbf{Coherence}} &
\makecell{\small\textbf{Pedagogical}\\\small\textbf{Effectiveness}} &
\makecell{\small\textbf{Typographic}\\\small\textbf{Clarity}} &
\makecell{\small\textbf{Diagram--Problem}\\\small\textbf{Alignment}} &
\makecell{\small\textbf{Element}\\\small\textbf{Layout Quality}} &
\makecell{\small\textbf{Visual}\\\small\textbf{Consistency}} &
\makecell{\small\textbf{Text--Diagram}\\\small\textbf{Coordination}} &
\small\textbf{Overall} & \makecell{\small\textbf{Success}\\\small\textbf{Rate}} \\
\midrule
Gemini 3.0 Pro Preview & \textbf{87.6\%} & \textbf{94.8\%} & \textbf{78.4\%} & \textbf{98.4\%} & \textbf{87.4\%} & \textbf{84.6\%} & 89.4\% & \textbf{92.6\%} & \textbf{87.8\%} & 97.4\% \\
Kimi-K2.5              & 85.2\% & 92.2\% & 73.4\% & 97.4\% & 71.4\% & 68.8\% & 88.6\% & 86.2\% & 80.8\% & \textbf{98.3\%} \\
Qwen3.5-397B           & \textbf{88.6\%} & 94.4\% & 70.2\% & 91.2\% & 49.2\% & 61.8\% & 83.6\% & 67.8\% & 72.0\% & 93.9\% \\
Qwen3.5-122B           & 83.4\% & 92.0\% & 71.2\% & 93.4\% & 42.8\% & 58.0\% & 84.6\% & 59.4\% & 68.6\% & 94.8\% \\
Qwen 3.5-35B           & 83.4\% & 89.0\% & 69.2\% & 87.8\% & 39.6\% & 53.6\% & 88.6\% & 55.0\% & 65.8\% & 83.9\% \\
GPT-5                  & 59.0\% & 66.6\% & 44.0\% & 74.2\% & 44.4\% & 59.2\% & \textbf{92.0\%} & 63.8\% & 58.0\% & 89.6\% \\
Claude Sonnet 4.5      & 55.2\% & 58.8\% & 47.6\% & 77.4\% & 42.6\% & 64.0\% & 84.2\% & 68.8\% & 57.8\% & 96.1\% \\
Mistral-Large-3        & 39.8\% & 40.0\% & 32.4\% & 82.0\% & 24.6\% & 47.6\% & 83.2\% & 45.0\% & 43.0\% & 74.8\% \\
Mistral-Small-4        & 39.8\% & 38.0\% & 30.2\% & 71.6\% & 23.8\% & 49.2\% & 81.6\% & 39.4\% & 40.8\% & 70.4\% \\
Ministral-3-14B        & 37.6\% & 37.0\% & 31.0\% & 77.6\% & 23.8\% & 49.6\% & \textbf{92.4\%} & 35.4\% & 41.0\% & 17.4\% \\
\bottomrule
\end{tabular}}
\caption{Model performance on the full EduIllustrate benchmark (n=230, 0--100\% scale). Overall: geometric mean of 8 dimensions. Success Rate: pipeline success rate. Bold indicates best per column. Per-subject and per-grade breakdowns are in Appendix~\ref{sec:full-results}.}
\label{tab:main-results}
\end{table*}

\begin{figure*}[t]
  \centering
  \includegraphics[width=\linewidth]{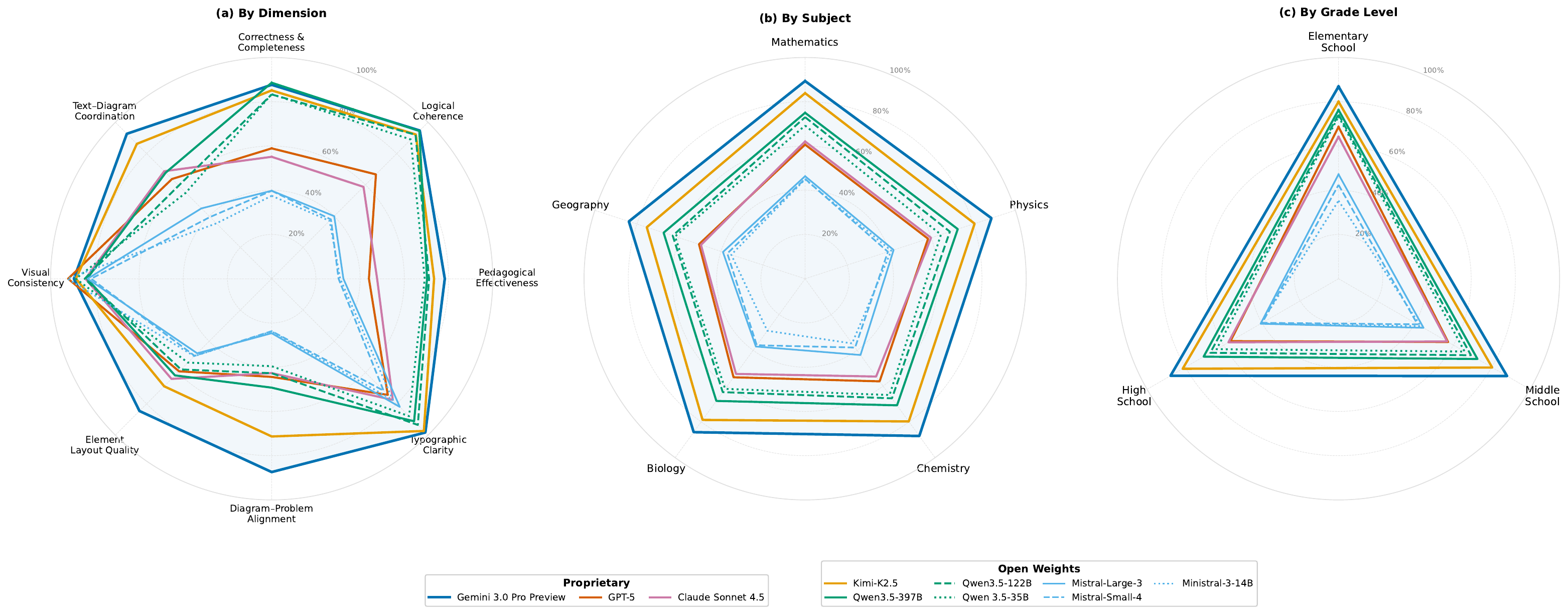}
  \caption{Radar charts of all 10 models. (a) By dimension: 8 evaluation dimensions. (b) By subject: overall scores on each of the five subjects. (c) By grade level: overall scores at each grade level.}
  \label{fig:radar}
\end{figure*}

Table~\ref{tab:main-results} presents overall performance across all eight dimensions. The overall score is computed as the geometric mean of all eight dimension scores, ensuring that a single critically low score substantially penalizes the overall result. Gemini 3.0 Pro Preview achieves the highest overall score (87.8\%), substantially outperforming all other models. Kimi-K2.5 ranks second (80.8\%). The Qwen family forms a middle tier (65.8\%--72.0\%), while GPT-5 and Claude Sonnet 4.5 score 58.0\% and 57.8\% respectively. The Mistral family scores lowest (40.8\%--43.0\%), with Ministral-3-14B severely limited by its 82.6\% failure rate. The over 46-point gap between the best (87.8\%) and worst (41.0\%) models demonstrates that architectural choices significantly impact multimodal educational content quality.

All models excel at Logical Coherence and Typographic Clarity, reflecting LLMs' strength in coherent, well-formatted text generation. Text--Diagram Coordination achieves high scores even for weaker models (Claude: 68.8\%, GPT-5: 92.0\%), indicating models produce well-integrated textual references to diagrams regardless of diagram quality. Notably, Ministral-3-14B and GPT-5 achieve the two highest Visual Consistency scores (92.4\% and 92.0\% respectively) despite weak overall performance; models with high failure rates tend to produce fewer scenes per problem, making cross-scene consistency trivially easier to maintain. Critical weaknesses appear in Diagram--Problem Alignment, where scores range from 23.8\% to 87.4\%. Qwen 3.5-35B's low score of 39.6\% and the Mistral family's scores near 23.8\%--24.6\% reflect frequent semantic misalignment. Pedagogical Effectiveness is universally weakest (30.2\%--78.4\%), indicating all models struggle with grade-appropriate language and scaffolding strategies. Failure rates on the 230-problem benchmark vary substantially: Kimi-K2.5 is most robust (1.7\%), followed by Gemini (2.6\%), Claude (3.9\%), Qwen3.5-122B (5.2\%), Qwen3.5-397B (6.1\%), GPT-5 (10.4\%), Qwen 3.5-35B (16.1\%), Mistral-Large-3 (25.2\%), Mistral-Small-4 (29.6\%), and Ministral-3-14B (82.6\%). The Mistral family---particularly Ministral-3-14B---shows substantially higher failure rates, indicating limited robustness to complex Manim code generation. Failed problems are excluded from scoring; reported scores reflect only successfully completed problems.

\begin{table*}[t]
\centering
\fontsize{10}{12}\selectfont
\setlength{\tabcolsep}{4pt}
\renewcommand\arraystretch{1.3}
\begin{tabular}{lcccccc|ccc}
\toprule
\textbf{Model} &
\makecell{\small\textbf{Mathe-}\\\small\textbf{matics}} &
\makecell{\small\textbf{Physics}} &
\makecell{\small\textbf{Chem-}\\\small\textbf{istry}} &
\makecell{\small\textbf{Biology}} &
\makecell{\small\textbf{Geog-}\\\small\textbf{raphy}} &
\makecell{\small\textbf{Subject}\\\small\textbf{Avg.}} &
\makecell{\small\textbf{Elem.}\\\small\textbf{School}} &
\makecell{\small\textbf{Middle}\\\small\textbf{School}} &
\makecell{\small\textbf{High}\\\small\textbf{School}} \\
\midrule
Gemini 3.0 Pro Preview & \textbf{89.4\%} & \textbf{88.6\%} & \textbf{87.9\%} & \textbf{85.7\%} & \textbf{83.8\%} & \textbf{87.1\%} & \textbf{86.9\%} & \textbf{88.0\%} & \textbf{87.8\%} \\
Kimi-K2.5              & 83.9\% & 80.6\% & 79.8\% & 78.9\% & 75.3\% & 79.7\% & 80.0\% & 80.3\% & 81.4\% \\
Qwen3.5-397B           & 75.0\% & 72.6\% & 70.7\% & 68.3\% & 67.3\% & 70.8\% & 76.3\% & 72.5\% & 70.5\% \\
Qwen3.5-122B           & 73.0\% & 68.9\% & 66.8\% & 63.4\% & 63.0\% & 67.0\% & 74.2\% & 69.2\% & 67.0\% \\
Qwen 3.5-35B           & 69.2\% & 64.6\% & 65.0\% & 61.5\% & 61.8\% & 64.4\% & 73.4\% & 65.8\% & 63.7\% \\
GPT-5                  & 60.6\% & 58.5\% & 57.3\% & 55.1\% & 50.5\% & 56.4\% & 68.7\% & 57.2\% & 56.4\% \\
Claude Sonnet 4.5      & 62.1\% & 59.9\% & 54.6\% & 53.2\% & 49.5\% & 55.9\% & 64.2\% & 56.7\% & 57.5\% \\
Mistral-Large-3        & 46.4\% & 42.1\% & 42.6\% & 37.9\% & 39.1\% & 41.6\% & 47.2\% & 44.4\% & 40.7\% \\
Mistral-Small-4        & 45.2\% & 39.4\% & 38.5\% & 37.2\% & 37.0\% & 39.5\% & 42.4\% & 41.3\% & 39.9\% \\
Ministral-3-14B        & 44.9\% & 41.0\% & 36.2\% & 29.1\% & 35.1\% & 37.3\% & 35.3\% & 42.9\% & 40.5\% \\
\bottomrule
\end{tabular}
\caption{Overall scores (geometric mean, 0--100\% scale) by subject and grade level for all 10 models. Subject Average: unweighted mean across five subjects. Bold indicates best per column.}
\label{tab:subject-grade}
\end{table*}

Per-subject and per-grade breakdowns with full dimension scores are provided in Appendix~\ref{sec:full-results}. Table~\ref{tab:subject-grade} summarizes the overall scores for five representative models across subjects and grade levels. Mathematics consistently achieves the highest scores across models, while geography scores lowest. Elementary school problems consistently yield higher scores than high school problems for most models (e.g., GPT-5: 68.7\% vs.\ 56.4\%; Claude: 64.2\% vs.\ 57.5\%), suggesting that increasing problem complexity and domain-specific reasoning demands degrade performance. Gemini and Kimi maintain relative consistency across grade levels, while other models show greater degradation on high school problems.

\subsection{Workflow Ablation Study}

\begin{figure}[t]
  \centering
  \includegraphics[width=\columnwidth]{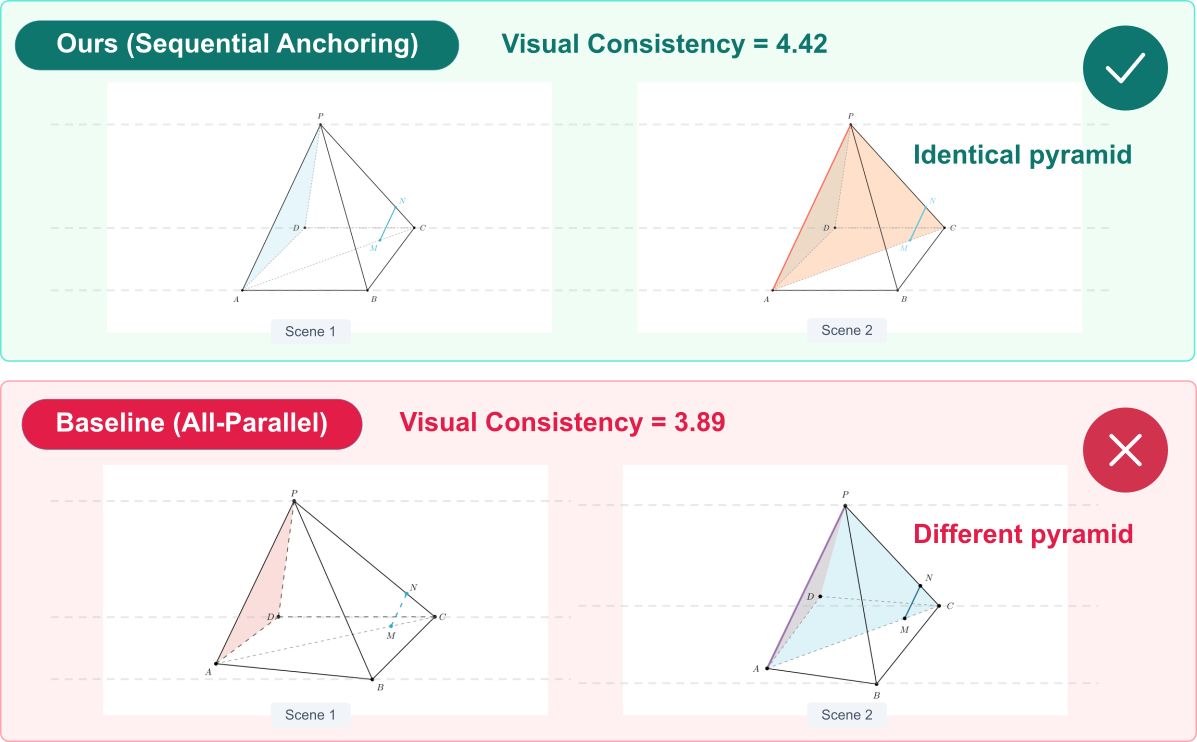}
  \caption{Visual consistency comparison between our sequential anchoring workflow (top) and the all-parallel baseline (bottom) on the same problem (four-pyramid $P$-$ABCD$). In our workflow, Scene 1 and Scene 2 render the identical pyramid structure---Scene 2 simply highlights the $PAC$ face on top of the same base diagram---maintaining full geometric and stylistic consistency. In the all-parallel baseline, the pyramid in Scene 1 and Scene 2 is drawn differently, breaking visual coherence and forcing the reader to reconcile two inconsistent representations of the same object.}
  \label{fig:case-study}
\end{figure}

EduIllustrate is built upon a modified version of the TheoremExplainAgent codebase~\citep{ku2025theoremexplain}, which originally adopts an all-parallel workflow where all scenes generate implementation plans and code in parallel. We compare our sequential Scene 1 + parallel Scenes 2-N workflow against this all-parallel baseline to quantify the impact of our anchoring strategy. The baseline generates and renders all scenes in parallel without Scene 1 conditioning. Table~\ref{tab:workflow} shows the all-parallel approach incurs 104\% higher input token consumption (95.5K vs. 46.8K)---because every scene independently generates a full implementation plan before code generation, whereas our workflow restricts planning to Scene 1 only---92\% higher cost (\$0.94 vs. \$0.49), and 13\% worse Visual Consistency (77.8\% vs. 89.4\%), while achieving lower overall quality (85.8\% vs. 87.8\%). Manual inspection reveals the all-parallel workflow's consistency failures stem from independent scene generation preventing style propagation. A side-by-side visual comparison is shown in Figure~\ref{fig:case-study}. This validates our sequential anchoring strategy.

\begin{figure}[t]
  \centering
  \includegraphics[width=0.75\columnwidth]{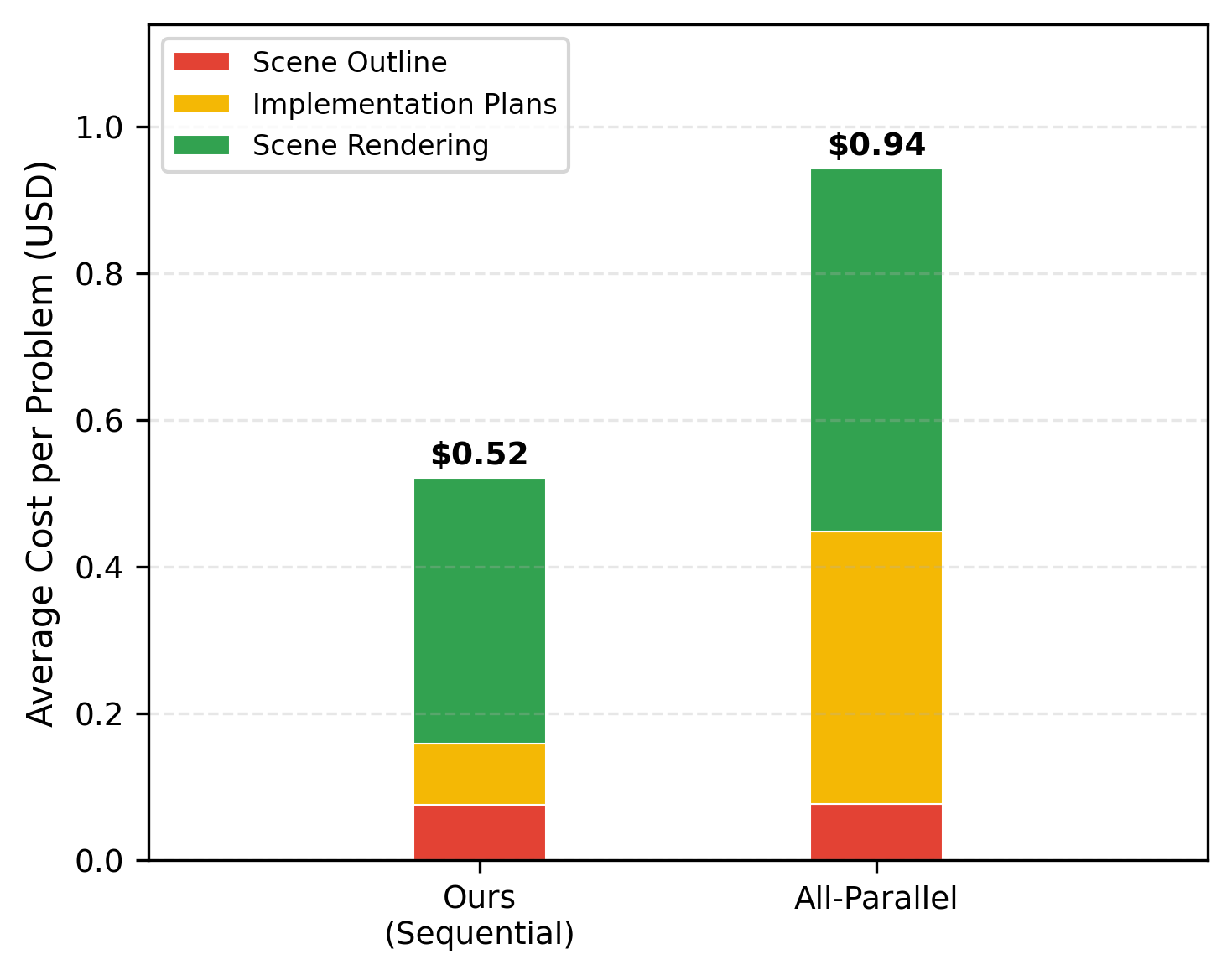}
  \caption{Per-stage cost comparison between our sequential anchoring workflow and the all-parallel baseline (both using Gemini 3.0 Pro Preview). The all-parallel approach incurs substantially higher Implementation Plans cost due to independent per-scene planning.}
  \label{fig:workflow-cost}
\end{figure}

\begin{table}[t]
\centering
\renewcommand\arraystretch{1.3}
\resizebox{\columnwidth}{!}{\begin{tabular}{lcccc}
\toprule
\makecell{\small\textbf{Workflow}} &
\makecell{\small\textbf{Tokens}\\\small\textbf{(In/Out)}} &
\makecell{\small\textbf{Cost}\\\small\textbf{(\$)}} &
\makecell{\small\textbf{Visual}\\\small\textbf{Consistency}} &
\small\textbf{Overall} \\
\midrule
Ours & 46.8K / 37.1K & 0.49 & 89.4\% & \textbf{87.8\%} \\
All-Parallel & 95.5K / 71.4K & 0.94 & 77.8\% & \textbf{85.8\%} \\
\bottomrule
\end{tabular}}
\caption{Workflow comparison using Gemini 3.0 Pro Preview on 50 problems. Our workflow achieves superior visual consistency at lower cost.}
\label{tab:workflow}
\end{table}

\subsection{Human-AI Agreement Analysis}

Table~\ref{tab:human-ai} presents inter-rater reliability (Krippendorff's $\alpha$) and human-AI agreement (Spearman's $\rho$) for 30 explanations evaluated across 7 dimensions by 20 expert raters. Logical Coherence and Diagram--Problem Alignment achieve strong human consensus ($\alpha \geq 0.80$) and strong human-AI agreement ($\rho \geq 0.80$), validating LLM-as-judge for these dimensions. Logical Coherence's high reliability reflects that logical gaps are objectively identifiable; Diagram--Problem Alignment's strong performance is notable given it requires visual reasoning. Typographic Clarity achieves strong-to-moderate agreement ($\rho = 0.77$, $\alpha = 0.67$), as formatting artifacts are largely objective but occasionally ambiguous at borderline cases. Pedagogical Effectiveness and Text--Diagram Coordination show moderate reliability ($\alpha \approx 0.64-0.68$, $\rho \approx 0.66-0.70$), acceptable for comparative studies but requiring human validation for high-stakes decisions. Visual Consistency and Element Layout Quality exhibit weak reliability ($\alpha \approx 0.30$, $\rho \approx 0.40$). Beyond ill-defined rubric constructs, a key contributing factor is that the LLM judge model is insensitive to low-level visual artifacts---such as overlapping elements, misaligned labels, or rendering glitches---that human raters readily detect when inspecting rendered diagrams.

\begin{table}[h]
\centering
\small
\begin{tabular}{lccl}
\hline
\textbf{Dim.} & \textbf{Spearman} & \textbf{Krippendorff} & \textbf{Reliability} \\
 & \textbf{$\rho$} & \textbf{$\alpha$} & \\
\hline
LC & 0.89 & 0.84 & Strong \\
DPA & 0.83 & 0.80 & Strong \\
TC & 0.77 & 0.67 & Strong-Mod. \\
PE & 0.70 & 0.68 & Moderate \\
TDC & 0.66 & 0.64 & Moderate \\
VC & 0.47 & 0.30 & Weak \\
ELQ & 0.39 & 0.32 & Weak \\
\hline
\end{tabular}
\caption{Human-AI agreement and inter-rater reliability (30 explanations, 20 raters). Correctness \& Completeness excluded (objective gold answer verification).}
\label{tab:human-ai}
\end{table}

\subsection{Cost Analysis}

\begin{table*}[t]
\centering
\small
\begin{tabular}{lccccc}
\hline
\textbf{Model} & \textbf{Avg Input} & \textbf{Avg Output} & \textbf{Input} & \textbf{Output} & \textbf{Avg Cost} \\
 & \textbf{Tokens} & \textbf{Tokens} & \textbf{(\$/M)} & \textbf{(\$/M)} & \textbf{(\$)} \\
\hline
Gemini 3.0 Pro Preview & 46,821 & 37,096 & 0.86 & 12.04 & 0.49 \\
Claude Sonnet 4.5 & 71,889 & 23,964 & 0.65 & 15.10 & 0.41 \\
Qwen3.5-122B-A10B & 39,487 & 29,262 & 0.40 & 3.20 & 0.11 \\
qwen3.5-397b & 39,937 & 27,285 & 0.60 & 3.60 & 0.12 \\
Kimi-K2.5 & 44,537 & 43,628 & 0.45 & 2.22 & 0.12 \\
GPT-5 & 34,568 & 10,178 & 0.74 & 10.00 & 0.13 \\
Mistral-Large-3 & 44,600 & 15,081 & 0.42 & 1.50 & 0.04 \\
Qwen 3.5-35B & 37,269 & 24,943 & 0.25 & 2.00 & 0.06 \\
Mistral-Small-4 & 54,195 & 17,833 & 0.13 & 0.60 & 0.02 \\
Ministral-3-14B & 50,488 & 18,128 & 0.17 & 0.20 & 0.01 \\
\hline
\end{tabular}
\caption{Cost analysis per problem (230 problems, average values).}
\label{tab:cost}
\end{table*}

\begin{figure}[t]
  \centering
  \includegraphics[width=\columnwidth]{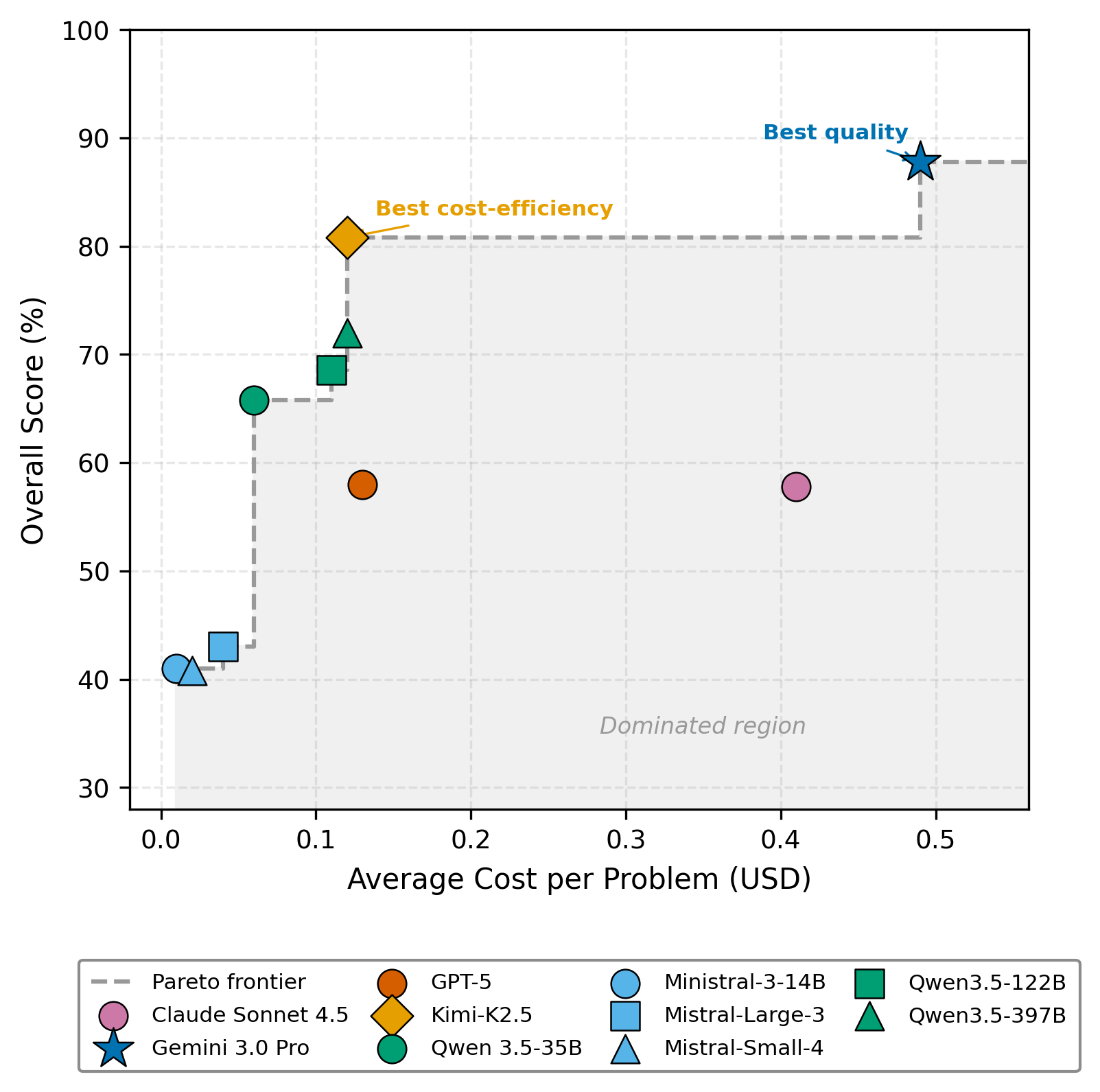}
  \caption{Cost vs.\ quality trade-off across ten models. X-axis: average cost per problem (USD); Y-axis: overall score (0--100\%). Kimi-K2.5 offers the best cost-efficiency among high-quality models, achieving 80.8\% at \$0.12/problem.}
  \label{fig:cost-quality}
\end{figure}

\begin{figure*}[t]
  \centering
  \includegraphics[width=\textwidth]{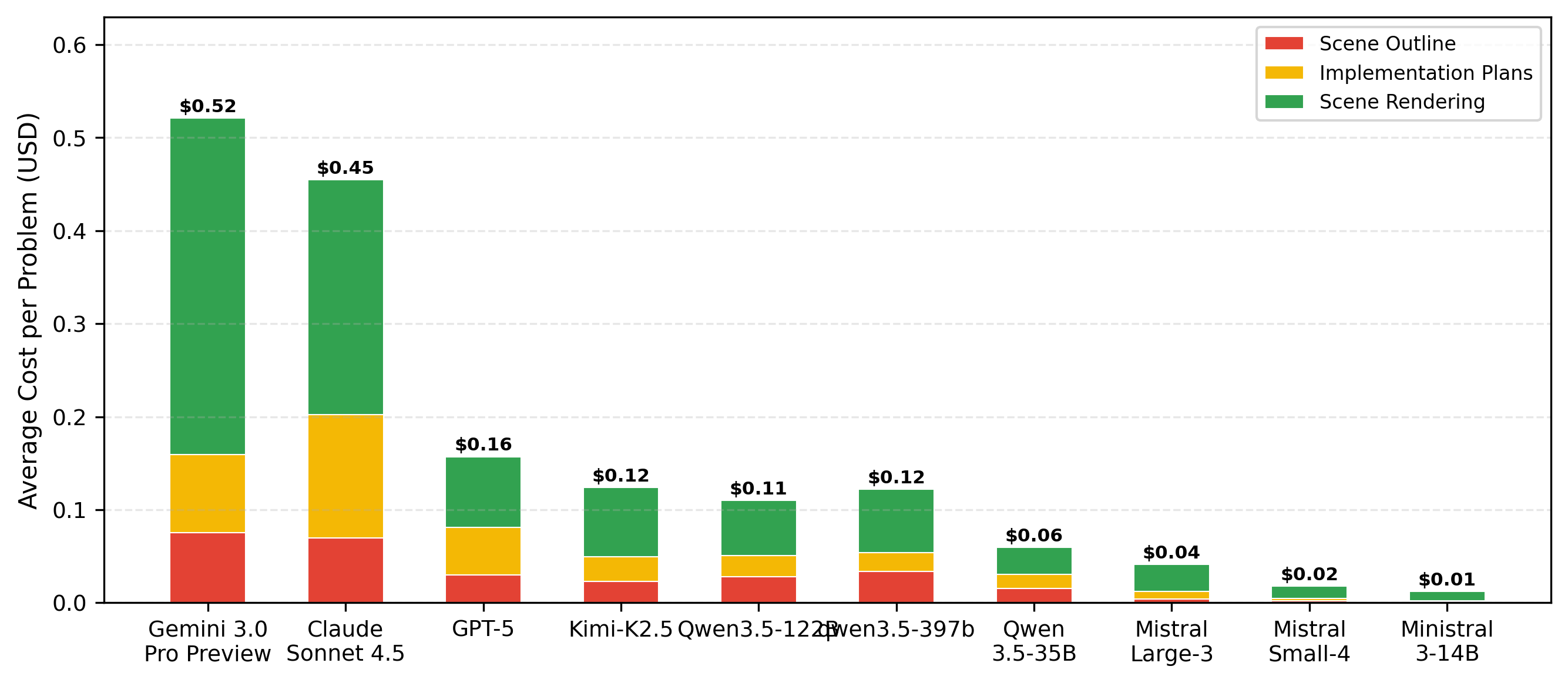}
  \caption{Per-stage cost breakdown across ten models. Each bar decomposes the average per-problem cost into three pipeline stages: Scene Outline, Implementation Plans, and Scene Rendering. Scene Rendering dominates cost for all models.}
  \label{fig:cost-breakdown}
\end{figure*}

Table~\ref{tab:cost} presents cost metrics across all ten models. As shown in Figure~\ref{fig:cost-quality}, Kimi-K2.5 offers the best cost-efficiency among high-quality models at \$0.12/problem (80.8\% score), representing only 8.0\% quality degradation relative to Gemini (\$0.49, 87.8\% score) at 4.1$\times$ lower cost. Among lower-cost options, Ministral-3-14B achieves the lowest cost at \$0.01/problem, while Mistral-Large-3 and Mistral-Small-4 offer competitive quality at \$0.04 and \$0.02 respectively.

\section{Conclusion}

We presented EduIllustrate, a benchmark for evaluating LLMs on K-12 STEM illustrated explanation generation, comprising a 230-problem dataset, a standardized generation protocol with sequential Scene 1 anchoring, and an 8-dimension rubric grounded in multimedia learning theory. Sequential anchoring improves Visual Consistency by 13\% over fully parallel generation at 94\% lower cost. Human evaluation validates LLM-as-judge reliability on objective dimensions while revealing limitations on subjective visual assessment.

\section*{Limitations}

The pipeline currently generates static PNG images via Manim, limiting both output modality and framework generalizability. While the agent framework can be adapted to produce animations by adjusting rendering flags and prompts, our 8-dimension rubric does not cover temporal coherence or animation pacing. The pipeline is also tightly coupled to Manim; extending to target-agnostic intermediate representations~\citep{kumar2025diagramir} would improve flexibility to other frameworks (TikZ, matplotlib, SVG).

Weak human-AI agreement on visual dimensions (Element Layout Quality: $\rho = 0.39$, Visual Consistency: $\rho = 0.47$) indicates current vision-language models struggle to assess layout quality and stylistic consistency reliably, and the low inter-rater agreement among humans ($\alpha \approx 0.30$) suggests these rubric constructs require further refinement. More broadly, Pedagogical Effectiveness is the weakest dimension across all models (50.8\%--78.4\%), as the pipeline delivers static, one-shot explanations with no adaptation to the individual learner. Future work will explore multi-turn Socratic interaction and student profile integration to personalize both explanation strategy and diagram design.

\section*{Acknowledgments}

We thank the 20 expert raters (15 STEM graduate students and 5 education doctoral students) for their meticulous human evaluation work.

\bibliography{custom}

\appendix

\section{Benchmark Construction Details}
\label{sec:benchmark-details}

\subsection*{Dataset Curation Process}

\begin{figure*}[!ht]
  \centering
  \includegraphics[width=\textwidth]{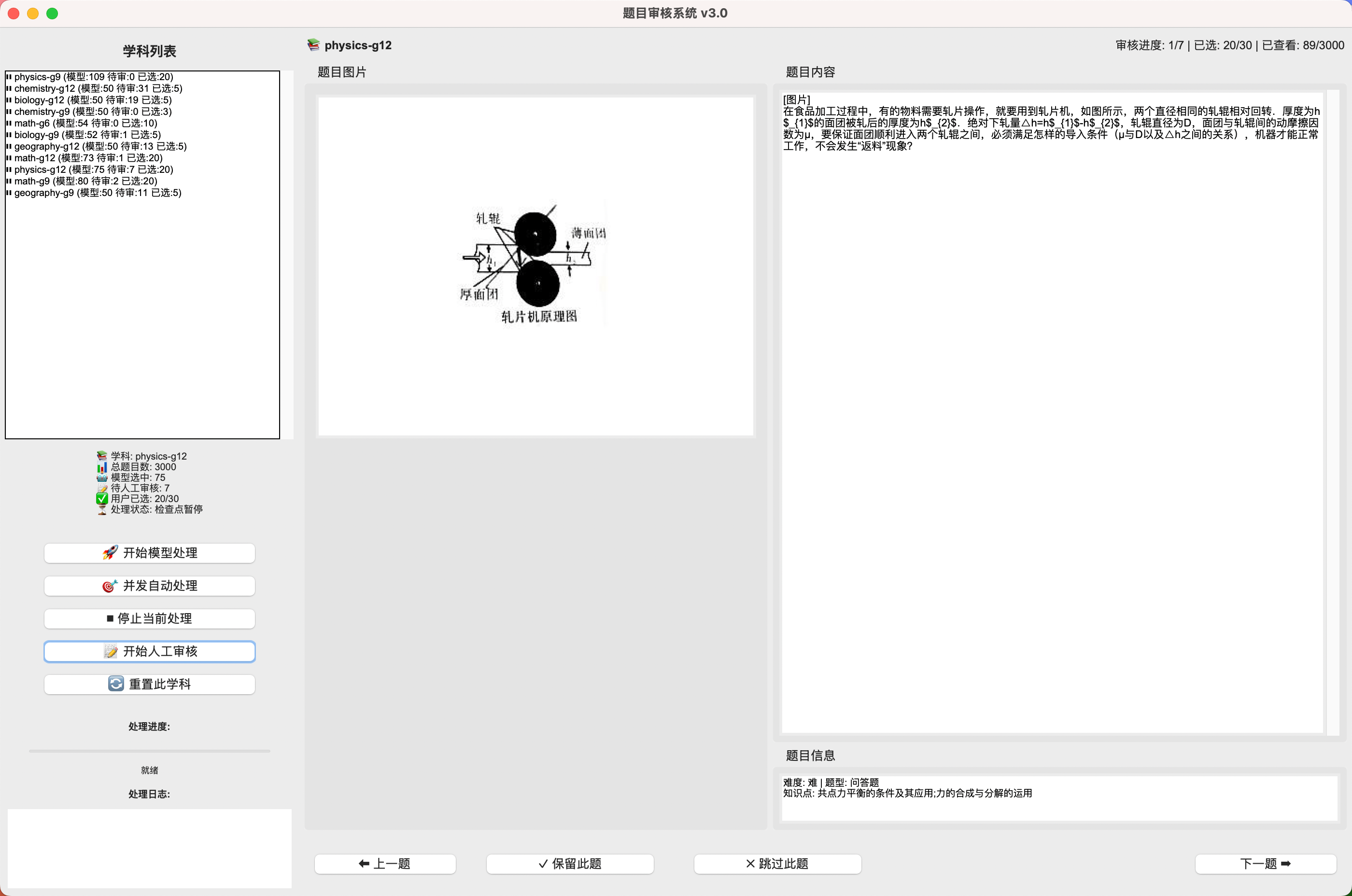}
  \caption{Screenshot of the problem review interface used during human annotation. Annotators evaluated each candidate problem against the three curation criteria and recorded their decisions with justifications.}
  \label{fig:question-reviewer}
\end{figure*}

Benchmark construction followed a two-phase procedure. In the first phase, Kimi-K2.5 automatically screened all candidate problems from K12-Vista~\citep{zhang2025k12vista} for \textbf{Diagram Appropriateness}: for each problem, the model was prompted to judge whether a visual representation would convey spatial, topological, or relational information essential to problem-solving (rather than merely restating the text), and to output a binary decision with a brief justification. In the second phase, two human annotators independently reviewed every problem against all three curation criteria---\textbf{(1) Diagram-Appropriate Problems}, \textbf{(2) Clear Gold Solutions}, and \textbf{(3) Diverse Topics}---and resolved disagreements through discussion, using a dedicated review interface (Figure~\ref{fig:question-reviewer}). Problems were retained only when both annotators confirmed all three criteria were satisfied. This two-phase design combines the scalability of LLM screening with the reliability of human expert judgment.

\subsection*{Topic Coverage}

Table~\ref{tab:benchmark-full} lists the complete topic coverage of the 230 benchmark problems, organized by grade level and subject.

\begin{table*}[!ht]
\centering
\small
\begin{tabular}{llp{10cm}}
\hline
\textbf{Level} & \textbf{Subject} & \textbf{Topics} \\
\hline
\multirow{5}{*}{High School}
 & Biology (15)
 & Law of independent assortment (×3); PCR amplification; Variants of 9:3:3:1 and 1:1:1:1 ratios (×2); Sex-linked inheritance (×3); Gene locus determination; Chromosomal structural variation; Genetic engineering; Transcription and translation; Ecosystem structure \\
\cline{2-3}
 & Chemistry (15)
 & Galvanic and electrolytic cell principles (×2); Crystal structure and properties; Organic compound structure and properties (×2); Chemical equilibrium and equivalent equilibrium (×2); Organic molecular formula determination; Structural features of carbon bonding; Isomer enumeration \\
\cline{2-3}
 & Geography (15)
 & Earth and maps; Latitude/longitude and coordinate grids (×2); Earth's revolution: noon solar altitude and day-length variation (×3); General characteristics of Earth's motion (×2); Geographic significance of Earth's motion; Air pressure belts and wind belts \\
\cline{2-3}
 & Mathematics (30)
 & Proportional segments and circles (×2); Hyperbola properties; Proposition truth and plane relations (×2); Inscribed polygon properties; Tangent line proof; Conic section common features; Plane-plane perpendicularity; Tangent-chord angle; Inductive reasoning; Parabola properties; Limits of sequences; Pyramid structure (×2); Ellipse properties; Law of sines; Sphere volume and surface area; Area/volume from three-view drawings; Similar triangles (×2); Spatial line-line relations; Arithmetic sequences; Spatial vectors and sphere; (some topics unlabeled) \\
\cline{2-3}
 & Physics (30)
 & Concurrent force equilibrium and electric field; Work-energy theorem with projectile motion; Work-energy theorem with charged particle in uniform field (×3); Work-energy theorem and mechanical energy conservation (×2); Conservation of momentum (×2); EMF from conductor cutting field lines and Joule's law (×3); Coulomb's law and Newton's second law; Mechanical energy conservation and projectile motion; Ideal gas law and enclosed gas pressure; SHM, vertical projection, and Hooke's law; Energy conservation, steady current, and force equilibrium; Velocity selector and charged particle in combined fields \\
\hline
\multirow{5}{*}{Middle School}
 & Biology (15)
 & DNA replication/transcription/translation calculations; Ecosystem and ecological system; Complete and incomplete metamorphosis; Gene–DNA–chromosome relationships; Gas exchange in lungs; Blood vessels (arteries, veins, capillaries); Classification of algae, mosses, and ferns; Urinary system and urine formation \\
\cline{2-3}
 & Chemistry (15)
 & Molecules, atoms, ions, elements and their relationships (×2); Molecular properties and acid-base indicators; Four basic reaction types and complex decomposition reactions; Oxygen production and properties; Chemical inference and reaction type identification; Solubility curves; Gas identification and purification; Particle model and conservation of mass (×2); Substance transformation; Substance identification and reaction type determination; Air composition measurement \\
\cline{2-3}
 & Geography (15)
 & Latitude/longitude and coordinate grids (×3); Earth's revolution and day-length variation; Earth's rotation; Day-night alternation and terminator; Seasonal formation; Direction and location using coordinate grids (×2) \\
\cline{2-3}
 & Mathematics (30)
 & Linear function applications; Triangle angle sum and exterior angles (×2); Triangle circumscribed circle; Quadratic/linear function graphs and coefficients (×2); Triangle congruence (SAS) and Pythagorean theorem; 3D solid nets (×2); Similar triangles (×2); Inequalities and linear/quadratic functions; 30° right triangle, inscribed angles, and diameter; Net folding; Proportional segments, midsegment, and 30° triangle; Rotation and area; Rational number operations; Square properties and coordinates; Similar triangles and equilateral triangle; Golden ratio; Fold transformations; Axial symmetry; Perpendicular bisector properties \\
\cline{2-3}
 & Physics (30)
 & Force balance and pulley systems; Light refraction ray diagrams; Convex lens imaging rules; Circuit fault diagnosis; Force and motion; Pressure and gravity vs.\ pressure; Pressure comparison and density; Balanced forces and energy changes; Spring force meter; Friction; Lever equilibrium (×2, including minimum force); Ohm's law and electromagnet (×2); Ohm's law applications (×2); Pulley rope tension, work, and power; Ammeter usage; Three circuit states; Dynamic circuit analysis (×3); Magnetic poles and Ampere's right-hand rule; Archimedes' principle (×2) \\
\hline
Elementary & Mathematics (20)
 & Three-view and net diagrams (×2); Observing 3D shapes from different directions; Shape composition; Perimeter of circles and rings; Position and direction (×3); Rotation and coordinate position; Perimeter of composite figures; Clever perimeter calculation; Inverse/direct proportion applications; Composite shape counting; Overlapping problems; Cube/cuboid nets and folding; Area comparison \\
\hline
\end{tabular}
\caption{Complete topic coverage of the 230 EduIllustrate benchmark problems. Numbers in parentheses after subject names indicate problem count; (×$n$) after a topic indicates $n$ problems on that topic. Some high school mathematics problems have unlabeled topics in the original K12-Vista dataset.}
\label{tab:benchmark-full}
\end{table*}

\section{Judge Model Self-Preference Analysis}
\label{sec:judge-bias}

A potential concern with using Gemini 3.0 Pro Preview as the judge model is self-preference bias: the judge may systematically inflate scores for outputs it generated. To investigate this, we sampled 20 problems and re-evaluated all outputs from all five models plus the all-parallel workflow variant (120 explanations total) using GPT-5 as an alternative judge, then compared the two scoring distributions.

\paragraph{Both Judges Exhibit Self-Preference, but Gemini's Is Smaller.} Table~\ref{tab:judge-bias} reports the mean overall score assigned by each judge. Both judges show self-preference: Gemini-as-judge scores its own outputs 9.2\% higher than GPT-5-as-judge does (88.8\% vs.\ 79.6\%), while GPT-5-as-judge scores its own outputs 20.6\% higher than Gemini-as-judge does (81.8\% vs.\ 61.2\%). Crucially, Gemini's self-preference magnitude ($|\Delta|{=}9.2\%$) is less than half of GPT-5's ($|\Delta|{=}20.6\%$). Moreover, GPT-5's self-preference is severe enough to alter rankings: GPT-5-as-judge ranks itself first, whereas Gemini-as-judge ranks it last. By contrast, Gemini-as-judge and GPT-5-as-judge agree on the \textbf{top-2 ranking} (Gemini $>$ Kimi), confirming that Gemini's self-preference does not distort the main comparative conclusions. Kimi-K2.5 serves as a neutral anchor with the smallest cross-judge gap ($|\Delta|{=}2.4\%$), further validating the evaluation framework's consistency on models with no judge affiliation.

\paragraph{Dimension-Level Agreement.} Table~\ref{tab:judge-dim} reports Pearson $r$, Spearman $\rho$, and MAE between the two judges across all 120 explanations for each dimension. Element Layout Quality ($r{=}0.58$, $\rho{=}0.56$) and Text--Diagram Coordination ($r{=}0.56$, $\rho{=}0.58$) show the highest agreement, reflecting relatively objective visual criteria. Diagram--Problem Alignment has the largest MAE (1.15), indicating the two judges disagree most on diagram--problem semantic matching. Typographic Clarity shows near-zero correlation ($r{=}0.10$, $\rho{=}0.12$, both non-significant), suggesting this dimension's assessment is highly judge-dependent. The overall score achieves moderate agreement ($r{=}0.50$, $\rho{=}0.43$, MAE${=}0.61$), indicating directionally consistent but numerically divergent scoring.

\paragraph{Implications.} Self-preference bias is present in both judge models, consistent with prior findings on LLM-as-judge self-preference~\citep{zheng2023judging}. We select Gemini over GPT-5 as the primary judge because (1) its self-preference is substantially smaller, (2) it does not alter the model ranking, and (3) the top-2 ranking is robust across both judges. Nevertheless, the moderate cross-judge agreement on overall scores (Spearman $\rho{=}0.43$) underscores that absolute score values should be interpreted with caution, and we recommend future work report multi-judge robustness checks.

\begin{table}[H]
\centering
\small
\begin{tabular}{lccr}
\hline
\textbf{Model (Generator)} & \textbf{Gemini} & \textbf{GPT-5} & \textbf{$\Delta$} \\
 & \textbf{Judge} & \textbf{Judge} & \\
\hline
Gemini 3.0 Pro Preview & 88.8\% & 79.6\% & $-$9.2\% \\
Kimi-K2.5              & 81.2\% & 78.8\% & $-$2.4\% \\
Qwen 3.5-35B           & 67.8\% & 74.6\% & +7.0\% \\
Claude Sonnet 4.5      & 63.4\% & 72.8\% & +9.4\% \\
GPT-5                  & 61.2\% & 81.8\% & +20.6\% \\
\hline
\end{tabular}
\caption{Mean overall scores (0--100\%) assigned by Gemini 3.0 Pro Preview and GPT-5 as judge models on the same 20 problems (120 explanations). $\Delta$ = GPT-5 judge $-$ Gemini judge. GPT-5-as-judge exhibits strong self-preference ($\Delta{=}+20.6\%$).}
\label{tab:judge-bias}
\end{table}

\begin{table}[H]
\centering
\small
\begin{tabular}{lccc}
\hline
\textbf{Dimension} & \textbf{Pearson} & \textbf{Spearman} & \textbf{MAE} \\
 & \textbf{$r$} & \textbf{$\rho$} & \\
\hline
ELQ & 0.58 & 0.56 & 0.62 \\
TDC & 0.56 & 0.58 & 0.74 \\
VC  & 0.48 & 0.46 & 0.45 \\
LC  & 0.45 & 0.38 & 0.74 \\
DPA & 0.40 & 0.42 & 1.15 \\
C\&C & 0.39 & 0.39 & 0.87 \\
PE  & 0.29 & 0.27 & 0.90 \\
TC  & 0.10$^{\dag}$ & 0.12$^{\dag}$ & 0.77 \\
Overall & 0.50 & 0.43 & 0.61 \\
\hline
\end{tabular}
\caption{Dimension-level agreement between Gemini and GPT-5 judges (120 explanations). $^{\dag}$Non-significant ($p > 0.05$). Dimensions sorted by Pearson $r$.}
\label{tab:judge-dim}
\end{table}

\section{A Gallery of Generated Explanations}
\label{sec:gallery}

We present representative high-quality and low-quality explanations generated by EduIllustrate to illustrate the range of outputs across different subjects and failure modes. Each entry shows the problem statement, key solution steps, and the corresponding rendered diagrams.

\subsection*{High-Quality Explanations}

\noindent \textbf{Middle School Physics --- Circuit Analysis.} This example demonstrates strong Visual Consistency and Text--Diagram Coordination. All three scenes use identical circuit drawing conventions, and each scene incrementally modifies the previous one---adding or removing a single wire---so that the student can track exactly what changed. The progressive structure mirrors effective pedagogical scaffolding.

\begin{examplebox}\scriptsize\noindent
\textbf{Problem:} The circuit is shown. Which analysis is incorrect? (A) After S is closed, the circuit short-circuits. (B) After S is closed, $L_1$ and $L_2$ are in parallel and both light up. (C) Removing wire $b$ makes $L_1$ and $L_2$ series-connected. (D) Moving wire $M$ from B to A makes $L_1$ and $L_2$ parallel.\\[4pt]
\textbf{Approach:} Trace current paths from the positive terminal to determine circuit topology under each scenario, then evaluate each option.\\[4pt]
\includegraphics[width=\linewidth]{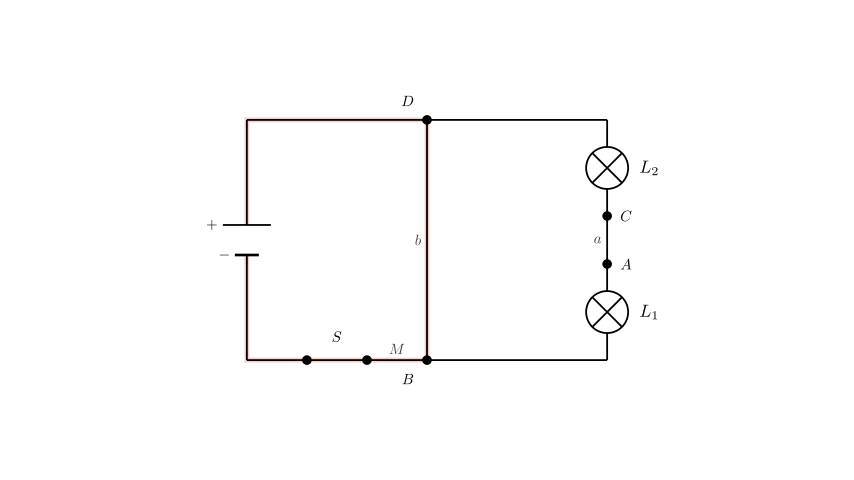}\\[2pt]
\scriptsize\textit{Scene 1: Original circuit diagram with current path analysis. Wire $b$ short-circuits the bulbs, creating a zero-resistance path.}

\textbf{Step 2:} Removing wire $b$ forces current through $L_2 \to L_1$ sequentially---series connection confirmed.\\[4pt]
\includegraphics[width=\linewidth]{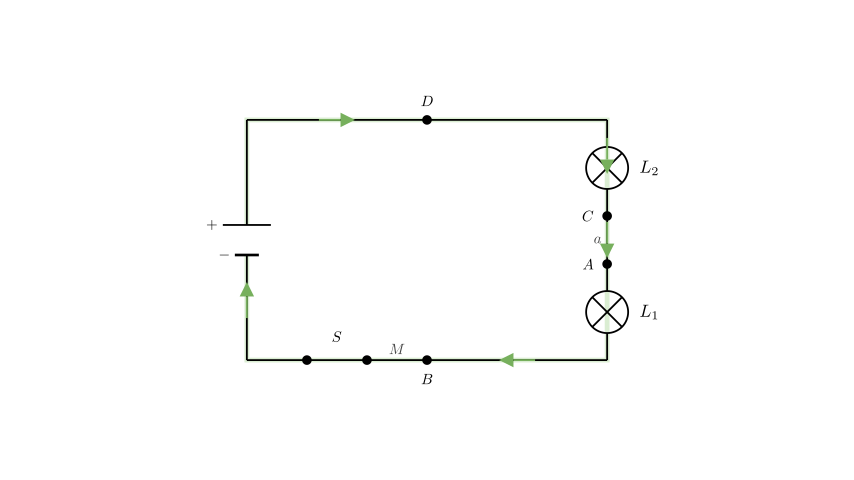}\\[2pt]
\scriptsize\textit{Scene 2: Modified circuit with wire $b$ removed. Single current path through both bulbs in series.}\\[4pt]
\textbf{Step 3:} Moving wire $M$ from B to A creates two independent paths. Both bulbs connect between the same high/low potential nodes---parallel connection.\\[4pt]
\includegraphics[width=\linewidth]{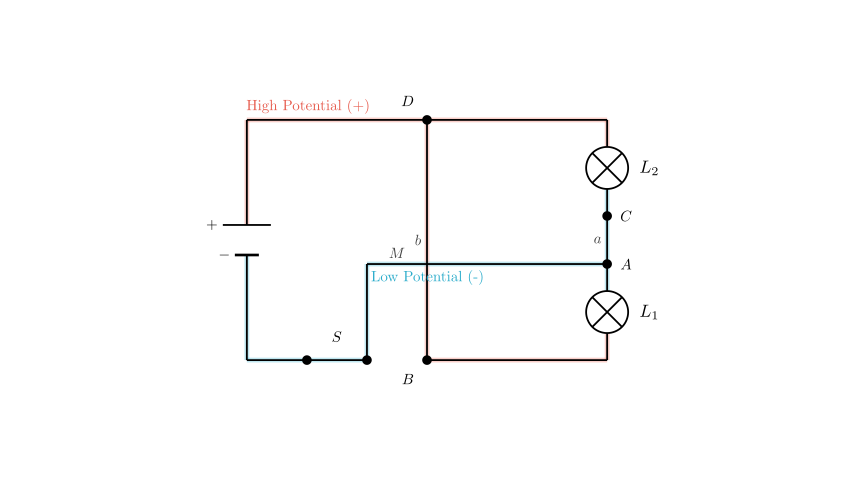}\\[2pt]
\scriptsize\textit{Scene 3: Circuit with wire $M$ moved to A. Both bulbs span identical potential difference---parallel.}\\[4pt]
\textbf{Answer: B} (bulbs do not light up; short circuit prevents current through loads).

\end{examplebox}

\noindent \textbf{Middle School Biology --- DNA Replication.} This example showcases consistent color coding (blue for $^{15}$N strands, orange for $^{14}$N strands) maintained across all three scenes. Each scene builds directly on the previous one, progressing from molecular structure to base composition to a replication tree. The color scheme serves as a visual thread that ties the explanation together---a hallmark of high Diagram--Problem Alignment.

\begin{examplebox}\scriptsize\noindent
\textbf{Problem:} A $^{15}$N-labeled eukaryotic gene has Cytosine (C) accounting for 30\% of all bases. Which statement is correct? (A) Helicase acts on sites \textcircled{1} and \textcircled{2}. (B) In one strand, (C+G)/(A+T) = 3:2. (C) After $T\to A$ mutation at site \textcircled{1}, after $n$ replications the mutant gene is 1/4. (D) After 3 replications in $^{14}$N medium, DNA with $^{14}$N is 3/4.\\[4pt]
\textbf{Approach:} Identify bond types at sites \textcircled{1} and \textcircled{2}; apply Chargaff's rules; apply semi-conservative replication.\\[4pt]
\includegraphics[width=\linewidth]{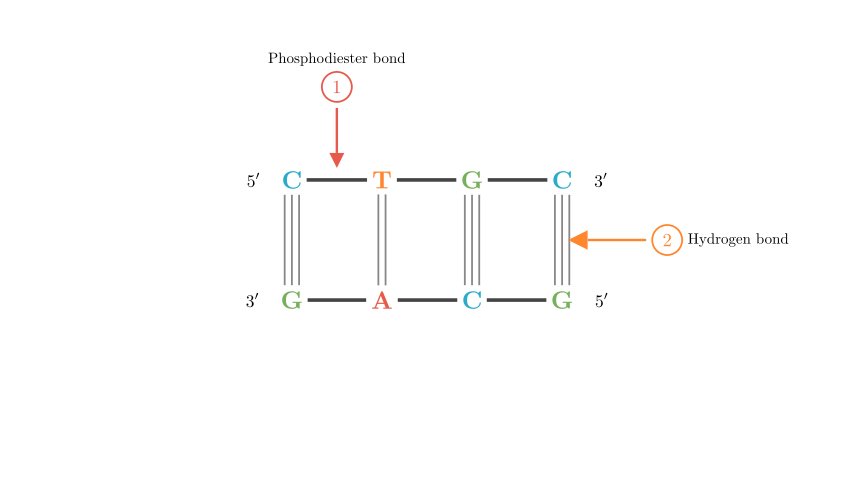}\\[2pt]
\scriptsize\textit{Scene 1: DNA double helix with sites \textcircled{1} (hydrogen bonds between bases) and \textcircled{2} (phosphodiester bonds in backbone) clearly labeled.}
\scriptsize
\textbf{Step 2:} Chargaff's rules: C=G=30\%, so A=T=20\%. Within one strand, C+G=60\%, A+T=40\%, giving ratio 3:2---Statement B is correct.\\[4pt]
\includegraphics[width=\linewidth]{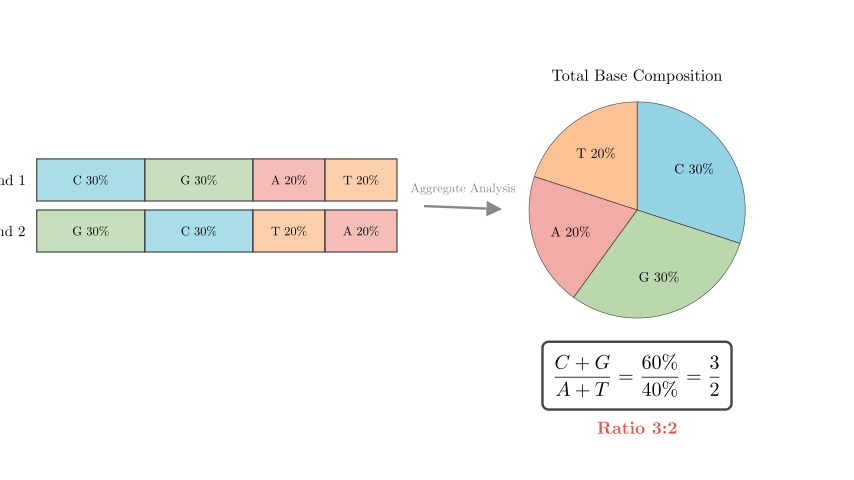}\\[2pt]
\scriptsize\textit{Scene 2: Base composition diagram showing C/G/A/T proportions derived from Chargaff's rules.}\\[4pt]
\textbf{Step 3:} Semi-conservative replication after 3 rounds yields $2^3=8$ DNA molecules; all 8 contain $^{14}$N (since new strands use $^{14}$N); 2 of 8 also retain $^{15}$N. So $^{14}$N proportion = 8/8 = 100\%, not 3/4---Statement D is wrong.\\[4pt]
\includegraphics[width=\linewidth]{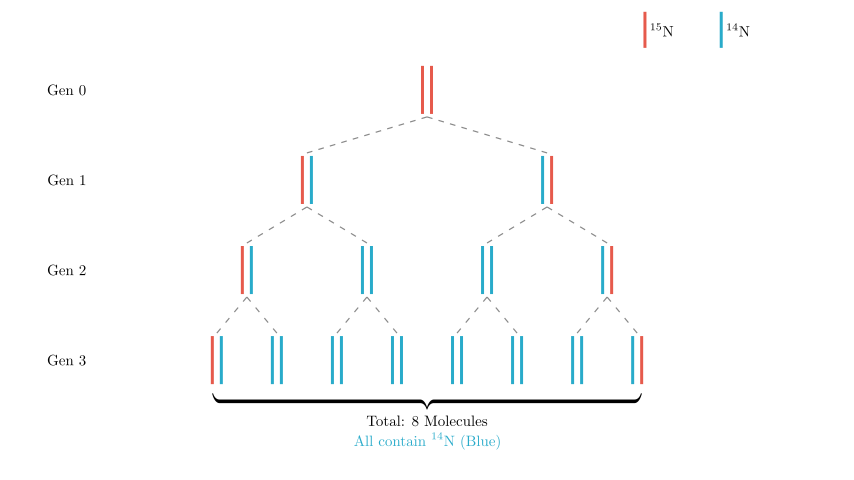}\\[2pt]
\scriptsize\textit{Scene 3: Semi-conservative replication tree after 3 rounds, showing isotope distribution across 8 daughter molecules.}\\[4pt]
\textbf{Answer: B.}

\end{examplebox}

\noindent \textbf{High School Math --- Geometric Series \& Inscribed Circles.} The three scenes progressively construct nested hexagons and circles, each building on the previous geometric structure. Annotations remain consistent (same label positions, line styles), and the final scene visualizes the infinite nesting---an abstract concept made concrete through visual accumulation. This illustrates how sequential anchoring preserves spatial coherence across increasingly complex diagrams.

\begin{examplebox}\scriptsize\noindent
\textbf{Problem:} A regular hexagon is inscribed in a circle of radius 1 m. Its inscribed circle is drawn, inside which another regular hexagon is inscribed, and so on infinitely. Find the sum $S$ of the areas of all circles.\\[4pt]
\textbf{Approach:} Find the ratio between successive circle radii using the apothem of a regular hexagon, identify the geometric series common ratio, then apply the infinite sum formula.\\[4pt]
\includegraphics[width=\linewidth]{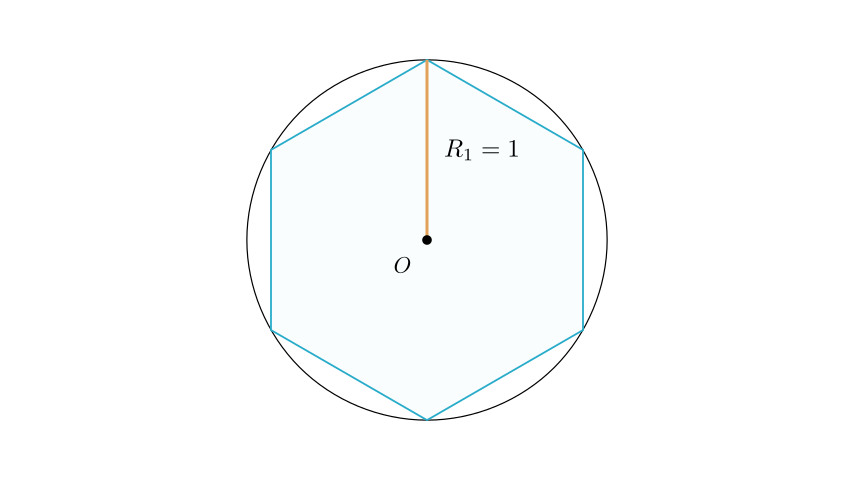}\\[2pt]
\scriptsize\textit{Scene 1: First circle ($R_1=1$) with inscribed hexagon and its apothem shown. The apothem equals $\frac{\sqrt{3}}{2}R_1$, giving $R_2=\frac{\sqrt{3}}{2}$.}
\scriptsize
\textbf{Step 2:} Area ratio $= \left(\frac{R_2}{R_1}\right)^2 = \frac{3}{4}$. The areas form a geometric series with first term $\pi$ and common ratio $\frac{3}{4}$.\\[4pt]
\includegraphics[width=\linewidth]{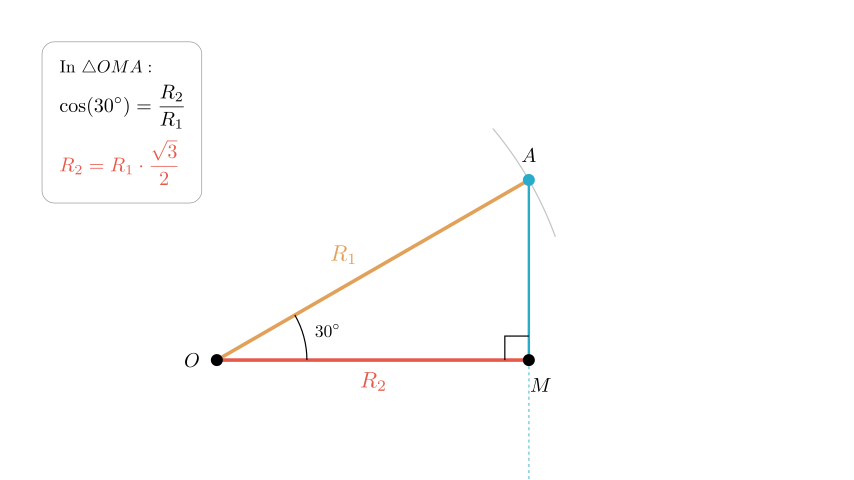}\\[2pt]
\scriptsize\textit{Scene 2: First two circles and hexagons overlaid, illustrating the scaling relationship.}\\[4pt]
\textbf{Step 3:} $S = \frac{\pi}{1 - \frac{3}{4}} = 4\pi$ m$^2$.\\[4pt]
\includegraphics[width=\linewidth]{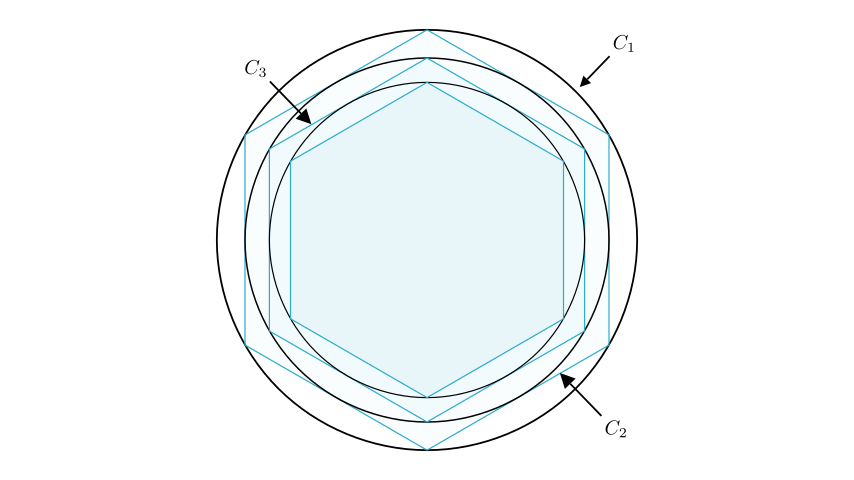}\\[2pt]
\scriptsize\textit{Scene 3: Infinite nesting visualized with converging circles and hexagons.}\\[4pt]
\textbf{Answer: $S = 4\pi$ m$^2$.}

\end{examplebox}

\noindent \textbf{High School Math --- Solid Geometry \& Perpendicular Planes.} This example requires multi-step spatial reasoning. The three scenes incrementally reveal the proof: Scene 1 establishes the rhombus base, Scene 2 proves $BD \perp$ plane $PAC$, and Scene 3 derives the perpendicularity condition. Each diagram preserves the same 3D viewpoint and labeling, allowing the reader to follow the logical chain without re-orienting spatially---strong evidence of both Visual Consistency and Logical Coherence.

\begin{examplebox}\scriptsize\noindent
\textbf{Problem:} In quadrangular pyramid $P$-$ABCD$, $PA \perp$ base $ABCD$, all base edges are equal (rhombus). $M$ is a moving point on $PC$. State a condition on $M$ such that plane $MBD \perp$ plane $PCD$.\\[4pt]
\textbf{Approach:} Use the plane-perpendicularity theorem: find a line in plane $MBD$ perpendicular to plane $PCD$. Show $BD \perp PC$ using the rhombus diagonal and $PA \perp$ base properties, then determine where $MBD$ must intersect $PCD$.\\[4pt]
\includegraphics[width=\linewidth]{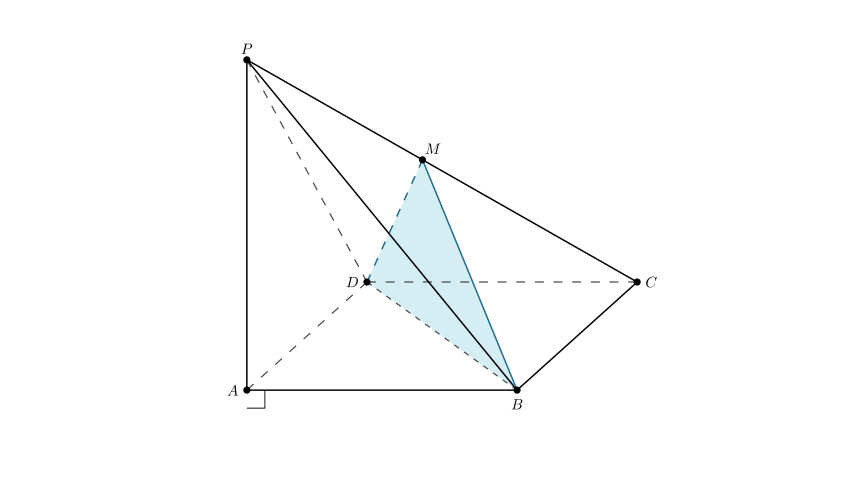}\\[2pt]
\textit{Scene 1: Pyramid with rhombus base $ABCD$, $PA$ perpendicular to base, diagonal $BD$ highlighted.}\\[6pt]
\textbf{Step 2:} Since $ABCD$ is a rhombus, $AC \perp BD$. Since $PA \perp$ base, $PA \perp BD$. Thus $BD \perp$ plane $PAC$, hence $BD \perp PC$.\\[4pt]
\includegraphics[width=\linewidth]{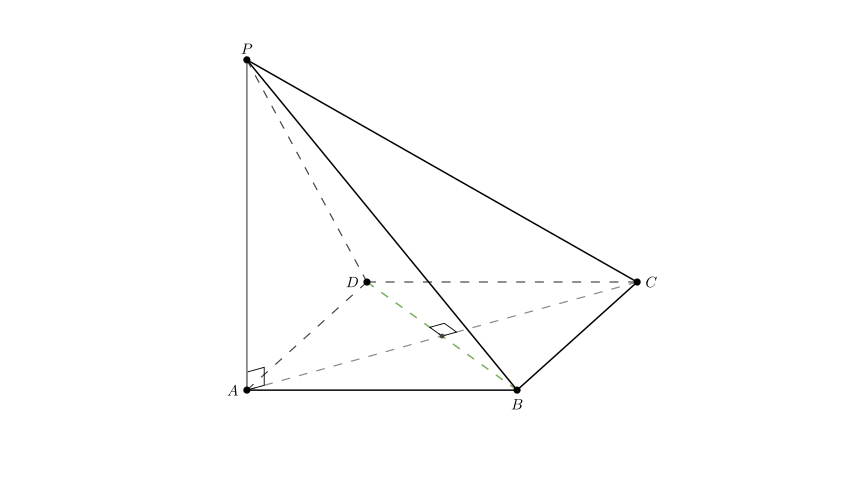}\\[2pt]
\textit{Scene 2: Plane $PAC$ highlighted, showing $BD \perp$ plane $PAC$ and therefore $BD \perp PC$.}\\[6pt]
\textbf{Step 3:} $BD \perp PC$ and $BD \subset$ plane $MBD$; if $M$ is the foot of the altitude from $B$ to $PC$ (i.e., $MB \perp PC$), then $BD \perp$ plane $PCD$, giving plane $MBD \perp$ plane $PCD$.\\[4pt]
\includegraphics[width=\linewidth]{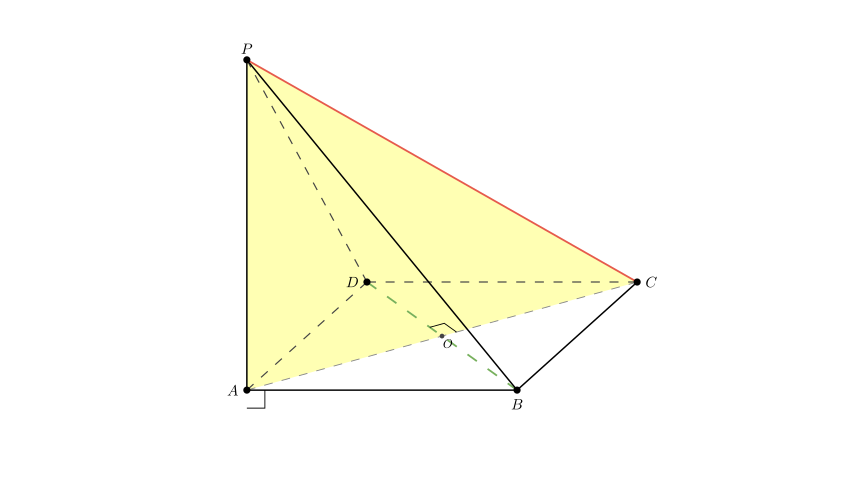}\\[2pt]
\textit{Scene 3: Plane $MBD$ and plane $PCD$ shown with their intersection line, confirming perpendicularity condition.}\\[4pt]
\textbf{Answer: $MB \perp PC$ (i.e., $M$ is the foot of the perpendicular from $B$ to $PC$).}

\end{examplebox}

\noindent \textbf{Middle School Math --- Quadratic Function Coefficients.} The two scenes effectively split the reasoning into visual subproblems: Scene 1 reads the signs of $a$ and $b$ from the parabola's shape, and Scene 2 plots the resulting linear function. Consistent axis styles and color coding across scenes reinforce the algebraic-to-graphical connection. The missing quadrant is clearly shaded, making the answer visually self-evident.

\begin{examplebox}\scriptsize\noindent
\textbf{Problem:} The graph of $y=ax^2+bx+1$ is shown. Which quadrant does $y=ax+b$ \textit{not} pass through?\\
A. First \quad B. Second \quad C. Third \quad D. Fourth\\[4pt]
\textbf{Approach:} Read signs of $a$ and $b$ from the parabola (opening direction, axis of symmetry position), then determine the linear function's slope and intercept to identify the missing quadrant.\\[4pt]
\includegraphics[width=\linewidth]{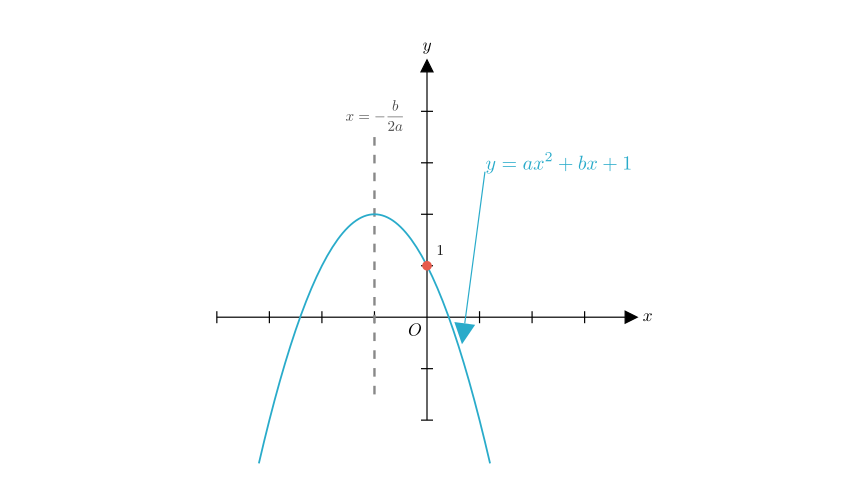}\\[2pt]
\scriptsize\textit{Scene 1: Parabola opens downward ($a<0$), axis of symmetry at $x<0$, so $-b/(2a)<0 \Rightarrow b<0$.}

\textbf{Step 2:} For $y=ax+b$: slope $a<0$ (decreasing), $y$-intercept $b<0$. A decreasing line with negative intercept passes through Quadrants II, III, IV---it does \textit{not} pass through Quadrant I.\\[4pt]
\includegraphics[width=\linewidth]{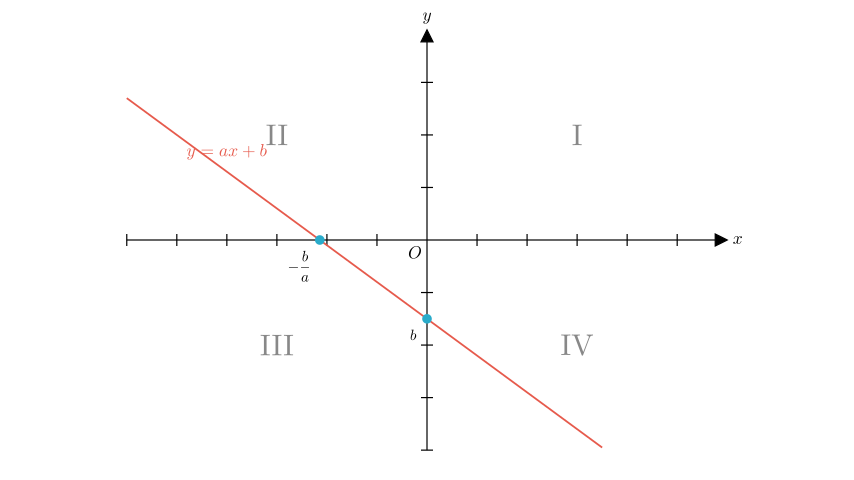}\\[2pt]
\scriptsize\textit{Scene 2: Linear function $y=ax+b$ plotted with $a<0$, $b<0$, passing through Q2, Q3, Q4 only.}\\[4pt]
\textbf{Answer: A (First quadrant).}

\end{examplebox}

\noindent \textbf{Middle School Math --- Square \& Rhombus Shaded Area.} A consistent color scheme (blue square, orange rhombus, green shaded region) is maintained across all three scenes, each progressively zooming into the region of interest. Scene 2 decomposes the unshaded area into computable triangles, and Scene 3 highlights the final answer. This example demonstrates effective use of color as a pedagogical anchor for the subtraction method.

\begin{examplebox}\scriptsize\noindent
\textbf{Problem:} Square $ABCD$ has area 25; rhombus $PQCB$ has area 20. Find the area of the shaded region.\\
A. 11 \quad B. 6.5 \quad C. 7 \quad D. 7.5\\[4pt]
\textbf{Approach:} Determine side lengths from given areas; use the subtraction method to express the shaded area as the square minus an unshaded triangular region formed by the overlap of the square and rhombus.\\[4pt]
\includegraphics[width=\linewidth]{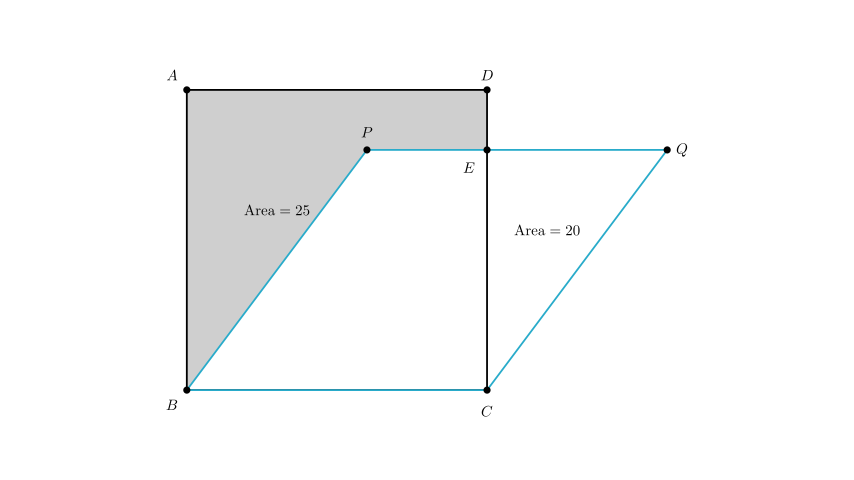}\\[2pt]
\scriptsize\textit{Scene 1: Square $ABCD$ (side=5) and rhombus $PQCB$ overlaid; the shaded region and the unshaded overlap region clearly distinguished by color.}

\textbf{Step 2:} Side of square = 5. Height of rhombus = area/base = 20/5 = 4. The unshaded region inside the square is a right triangle with legs 5 and $5-4=1$; its area = $\frac{1}{2}\times5\times1=2.5$. Shaded area $= 25 - 2\times2.5 \times \ldots$ (subtraction applied per configuration).\\[4pt]
\includegraphics[width=\linewidth]{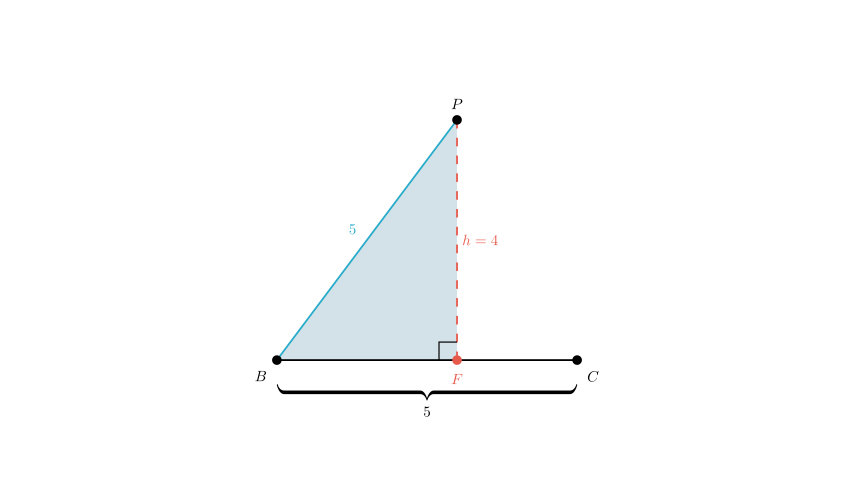}\\[2pt]
\scriptsize\textit{Scene 2: Decomposition of the unshaded region into computable triangles.}\\[4pt]
\includegraphics[width=\linewidth]{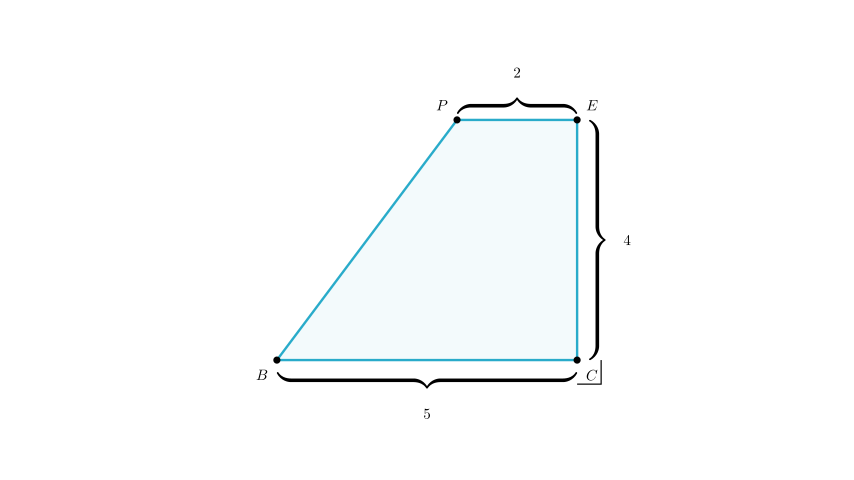}\\[2pt]
\scriptsize\textit{Scene 3: Final shaded region highlighted with computed area.}\\[4pt]
\textbf{Answer: D (7.5).}

\end{examplebox}

\subsection*{Low-Quality Explanations}

\noindent \textbf{Elementary Math --- Cube Net Folding (Correctness Failure).} The model applies the opposite-face pairing rule correctly in principle but misidentifies the symbol placement during the mapping step, arriving at answer C instead of the gold answer A. This is a pure Correctness \& Completeness failure: the textual reasoning framework is sound, but a single perceptual error in reading the net topology propagates to an incorrect conclusion. Such failures highlight the gap between procedural competence and visual perception accuracy.

\begin{examplebox}\scriptsize\noindent
\textbf{Problem:} Xiaoxiao made a cubic gift box where opposite faces share the same pattern ($\star$, $\heartsuit$, $\odot$). Which net (A/B/C/D) folds into this cube?\\[4pt]
\textbf{Approach:} Apply opposite-face pairing rule for cross-type nets: positions (1,3), (2,4), and (top, bottom) are opposite pairs.\\[4pt]
\includegraphics[width=\linewidth]{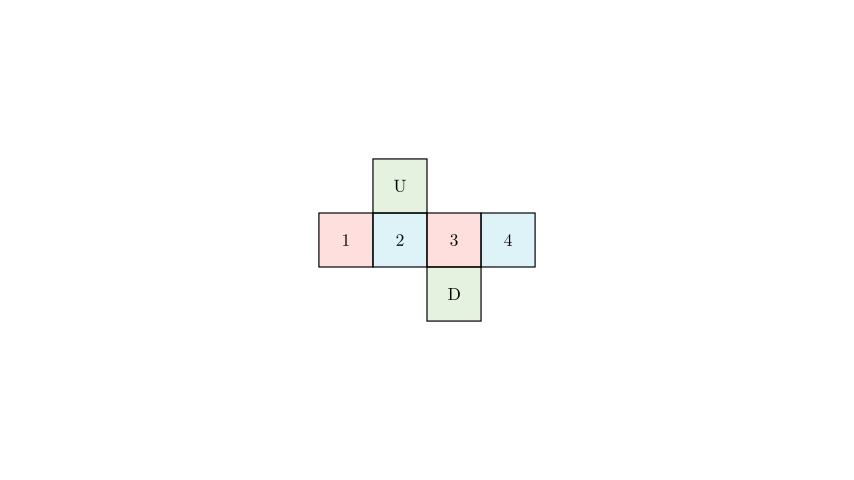}\\[2pt]
\scriptsize\textit{Scene 1: Cross-type net with positions labeled and opposite-pair rule illustrated.}

\textbf{Step 2:} Tests each option by mapping symbols to opposite-pair positions. Concludes that option C satisfies the constraint.\\[4pt]
\includegraphics[width=\linewidth]{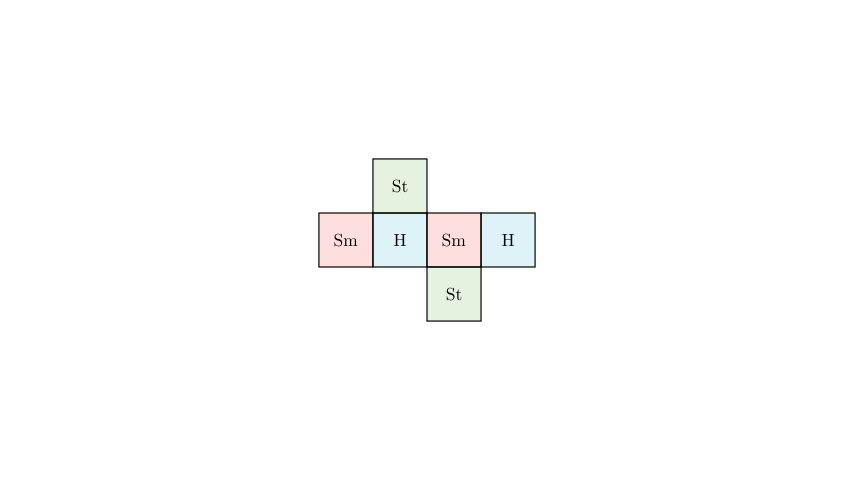}\\[2pt]
\scriptsize\textit{Scene 2: Net options annotated with opposite-face verification.}\\[6pt]
\textbf{Final Answer: C} \hspace{4pt} \textcolor{red}{$\times$ \textbf{Gold answer: A}}\\[4pt]
\textbf{Failure mode (Correctness \& Completeness):} The reasoning process applies the opposite-face pairing rule correctly in principle, but misidentifies the symbol placement in option A vs.\ C during the mapping step, leading to an incorrect final answer.

\end{examplebox}

\noindent \textbf{Middle School Math --- Isosceles Triangle (Diagram--Problem Alignment Failure).} Although the textual reasoning is correct ($\cos 36^\circ = \frac{\sqrt{5}+1}{4}$), the rendered diagrams draw an acute scalene triangle with visibly unequal sides, directly violating the given $AB=AC$ constraint. The $72^\circ$ base angles are not rendered at vertices B and C. A student relying on the diagrams would form incorrect spatial intuitions about the triangle's shape, undermining the pedagogical value despite textual correctness.

\begin{examplebox}\scriptsize\noindent
\textbf{Problem:} In $\triangle ABC$, $AB=AC=4$, $\angle C=72^\circ$. Point $D$ is the midpoint of $AB$; point $E$ lies on $AC$; $DE \perp AB$. Find $\cos A$.\\
Options: A. $\frac{\sqrt{5}-1}{2}$ \quad B. $\frac{\sqrt{5}-1}{4}$ \quad C. $\frac{\sqrt{5}+1}{4}$ \quad D. $\frac{\sqrt{5}+1}{2}$\\[4pt]
\textbf{Approach:} Use isosceles triangle properties ($AB=AC \Rightarrow \angle B=\angle C=72^\circ$), compute $\angle A = 180^\circ-144^\circ=36^\circ$, then evaluate $\cos 36^\circ$.\\[4pt]
\includegraphics[width=\linewidth]{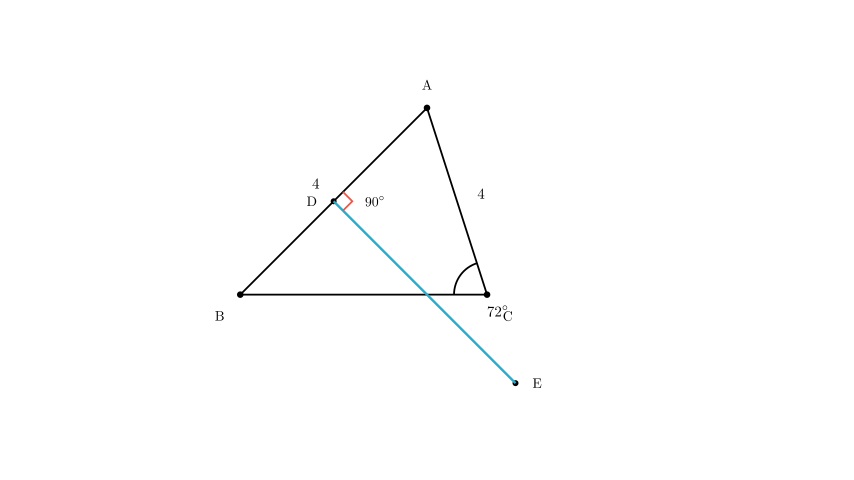}\\[2pt]
\scriptsize\textit{Scene 1: \textcolor{red}{The diagram draws an acute scalene triangle with visibly unequal sides, violating the given $AB=AC$ constraint. The $72^\circ$ angles are not rendered at vertices B and C.}}
\scriptsize
\textbf{Step 2:} $\angle A + 72^\circ + 72^\circ = 180^\circ \Rightarrow \angle A = 36^\circ$.\\[4pt]
\includegraphics[width=\linewidth]{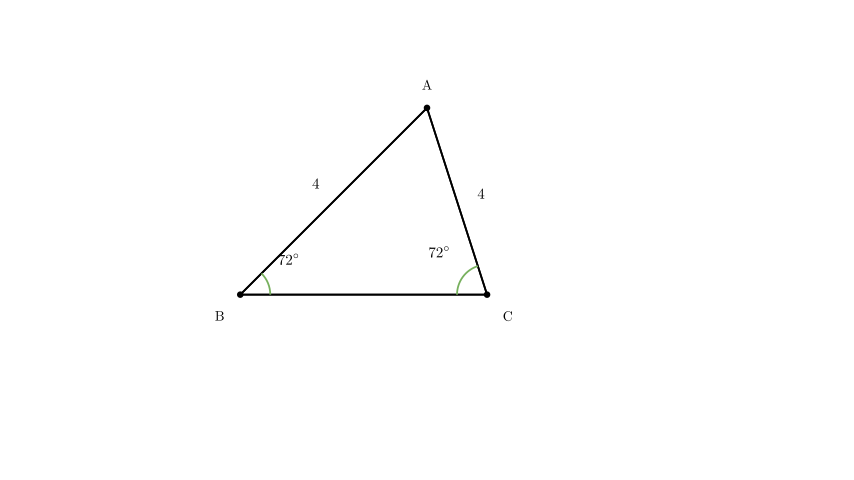}\\[2pt]
\scriptsize\textit{Scene 2: \textcolor{red}{The highlighted triangle still shows inconsistent proportions. The isosceles constraint ($AB=AC$) is visually absent, making the diagram misleading.}}\\[4pt]
\textbf{Step 3:} $\cos 36^\circ = \frac{\sqrt{5}+1}{4}$\\[4pt]
\textbf{Final Answer: C} \hspace{4pt} \textcolor{green!60!black}{$\checkmark$} (textually correct)\\[4pt]
\textbf{Failure mode (Diagram--Problem Alignment):} Although the textual reasoning is correct, the rendered diagrams do not reflect the isosceles structure of the triangle---a fundamental geometric constraint of the problem. A student relying on the diagrams would form incorrect spatial intuitions.

\end{examplebox}

\noindent \textbf{High School Physics --- Gas Laws (Element Layout \& Alignment Failure).} A critical geometric parameter ($3H/4$, the nitrogen column height in cylinder A) is mislabeled as $H$ in Scene 1, and this error propagates into Scene 2. The textual solution computes correct volumes using $3H/4$, but a student using the diagram to set up equations would obtain wrong initial volumes. This exemplifies how Element Layout Quality failures can actively mislead learners even when the text is correct.

\begin{examplebox}\scriptsize\noindent
\textbf{Problem:} Two connected vertical cylinders A (closed top, diameter $2d_B$, cross-section $4S$) and B (open top, diameter $d_B$, cross-section $S$), both height $H$. Piston $a$ in A is $H/4$ from the top; piston $b$ in B is at mid-height. $N_2$ fills below both pistons; $O_2$ fills above piston $a$. (1) Heat $N_2$ until piston $b$ reaches the top: find $N_2$ temperature. (2) Continue heating until piston $a$ rises $H/16$: find $O_2$ pressure.\\[4pt]
\textbf{Approach:} Stage I: $N_2$ expands at constant pressure $P_0$ (piston $b$ free); $O_2$ isothermal (constant volume). Stage II: $N_2$ volume in B fixed; $O_2$ compressed isothermally by Boyle's law.\\[4pt]
\includegraphics[width=\linewidth]{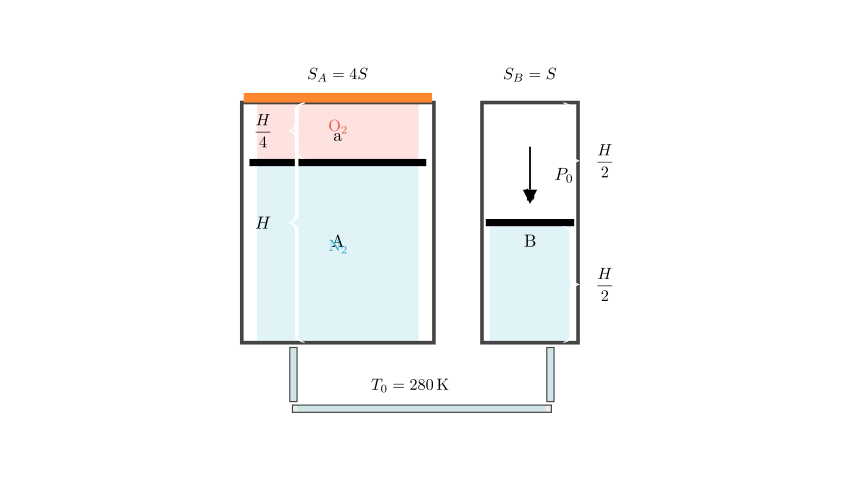}\\[2pt]
\scriptsize\textit{Scene 1: \textcolor{red}{The height label below piston $a$ in cylinder A is marked $H$ instead of the correct $3H/4$, contradicting the given initial geometry and making the diagram inconsistent with the problem.}}
\scriptsize
\textbf{Step 2:} Stage I---$N_2$ volume changes from $V_{N0}=4S\cdot\frac{3H}{4}+S\cdot\frac{H}{2}=\frac{7SH}{2}$ to $V_{N1}=\frac{3SH}{4}\cdot4S+SH=4SH$. By $pV\propto T$: $T_1=T_0\cdot\frac{4SH}{7SH/2}=\frac{8}{7}T_0$.\\[4pt]
\includegraphics[width=\linewidth]{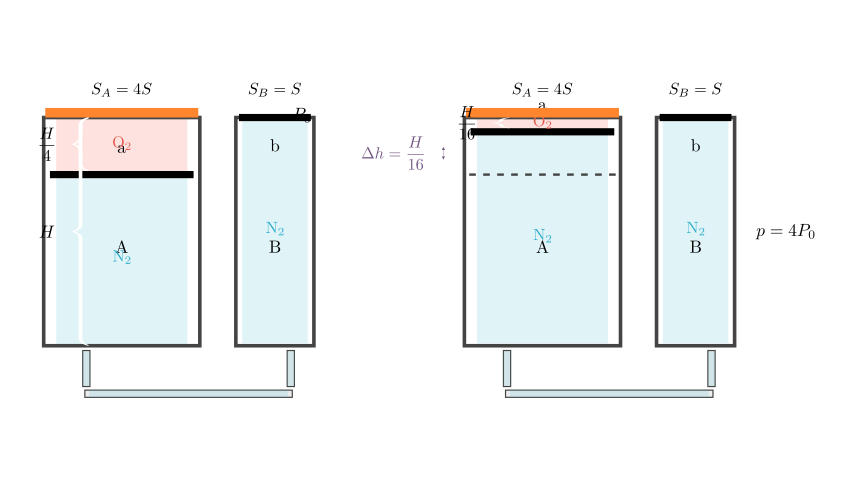}\\[2pt]
\scriptsize\textit{Scene 2: \textcolor{red}{After piston $b$ reaches the top, the diagram still carries the incorrect label from Scene 1, propagating the error into Stage II.}}\\[4pt]
\includegraphics[width=\linewidth]{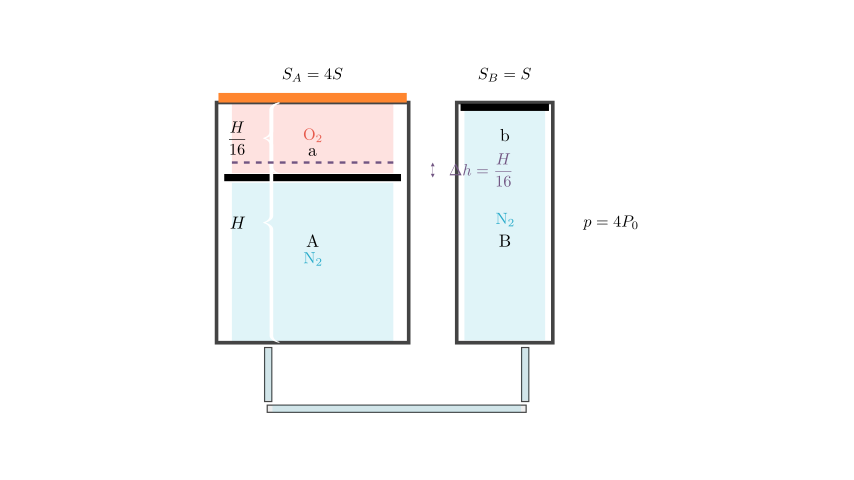}\\[2pt]
\scriptsize\textit{Scene 3: Stage II diagram showing piston $a$ risen by $H/16$; layout partially corrected but inconsistent with Scene 1.}\\[4pt]
\textbf{Failure mode (Element Layout Quality / Diagram--Problem Alignment):} The mislabeled height $H$ (should be $3H/4$) in Scene 1 violates the problem's geometric constraints. A student using the diagram to set up equations would obtain wrong initial volumes and incorrect answers.

\end{examplebox}

\noindent \textbf{Middle School Geography --- Latitude/Longitude Reading (Logical Coherence Failure).} The model arrives at the correct answer (B) through systematically flawed reasoning: it misidentifies point A's latitude zone and falsely claims the diagram lacks longitude reference markers. This ``right answer, wrong method'' failure is pedagogically dangerous---if adopted by students, it instills incorrect map-reading habits. The case reveals that Correctness \& Completeness alone is insufficient to assess educational value; Logical Coherence and Diagram--Problem Alignment must be evaluated jointly.

\begin{examplebox}\scriptsize\noindent
\textbf{Problem:} From a polar-view diagram of Earth with latitude circles and longitude markings, determine which statement is correct: (A) Point A receives direct sunlight only once a year. (B) Point B may receive direct sunlight twice a year. (C) Point B is in the Southern Hemisphere. (D) A is in the Eastern Hemisphere, B is in the Western Hemisphere.\\[4pt]
\textbf{Key facts:} Direct sunlight occurs only between $23.5^\circ$S and $23.5^\circ$N. Points on the tropics receive it once; points inside receive it twice. Hemisphere boundaries: equator (N/S), 20$^\circ$W--160$^\circ$E (E/W).\\[4pt]
\includegraphics[width=\linewidth]{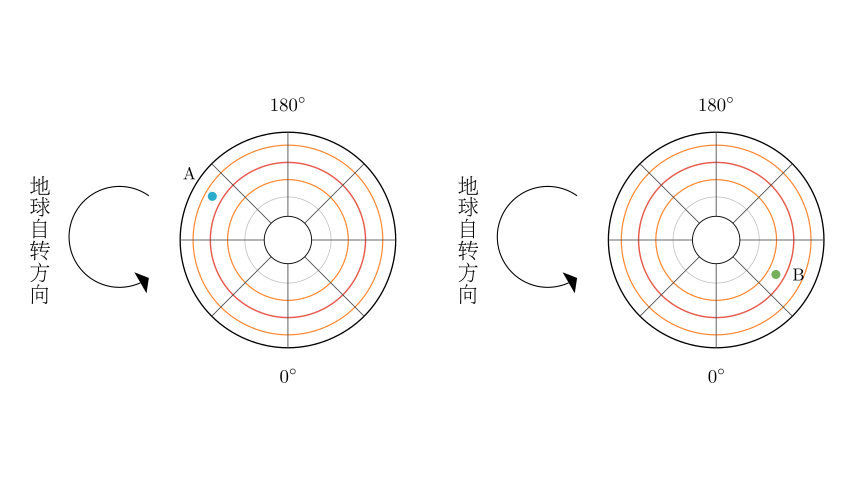}\\[2pt]
\scriptsize\textit{Scene 1: Polar-view diagram. \textcolor{red}{The model incorrectly identifies point A as lying outside the tropics (mid-latitude), concluding A receives no direct sunlight---in fact, A is on the equator and receives direct sunlight twice a year.}}
\scriptsize
\textbf{Flawed reasoning:} The model misreads A as a mid-latitude point, eliminating option A with wrong logic. It then claims the diagram lacks explicit $0^\circ$/$180^\circ$ longitude markings to determine the E/W hemisphere---but the diagram does contain computable longitude references, making option D verifiable. The model eliminates options A, C, D through incorrect intermediate steps and selects B by elimination.\\[4pt]
\includegraphics[width=\linewidth]{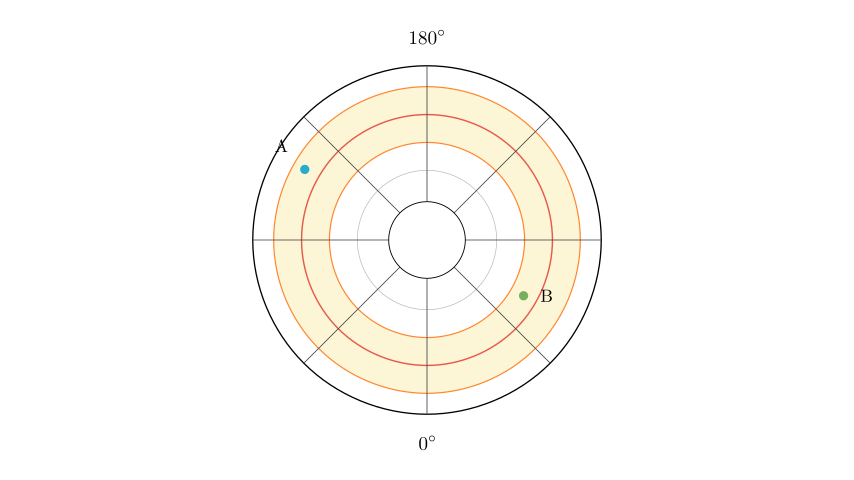}\\[2pt]
\scriptsize\textit{Scene 2: Hemisphere analysis diagram. \textcolor{red}{The longitude reference lines drawn are inconsistent with the original diagram's markings, reflecting the model's failure to correctly read the given coordinate information.}}\\[4pt]
\textbf{Final Answer: B} \hspace{4pt} \textcolor{green!60!black}{$\checkmark$} (coincidentally correct)\\[4pt]
\textbf{Failure mode (Logical Coherence / Diagram--Problem Alignment):} Correct final answer reached through systematically flawed reasoning---wrong latitude identification for A, and incorrect claim about missing longitude markers. This is a ``right answer, wrong method'' failure that would mislead students.

\end{examplebox}

\section{Human Annotation Process}
\label{sec:annotation}
\begin{figure*}[!ht]
  \centering
  \includegraphics[width=\textwidth]{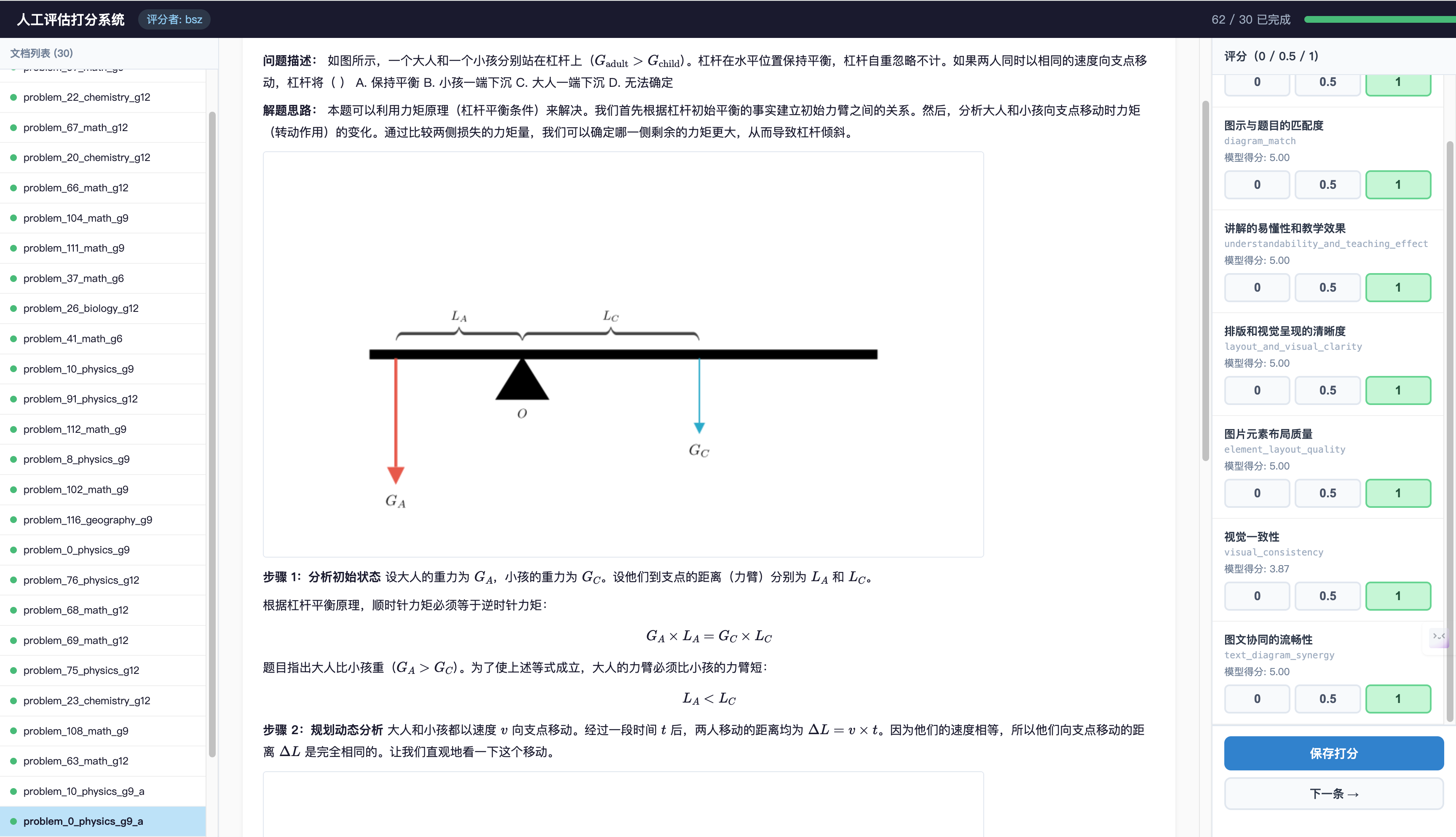}
  \caption{Screenshot of the human annotation website used for expert evaluation. Raters scored 7 dimensions on a 3-level scale after reviewing the problem, gold solution, and generated explanation with diagrams.}
  \label{fig:annotation}
\end{figure*}

Human evaluation was conducted through a dedicated annotation website (Figure~\ref{fig:annotation}). Raters were presented with a K-12 STEM problem, its gold-standard solution, and the generated illustrated explanation (including rendered diagrams). For each explanation, raters scored 7 dimensions (Logical Coherence through Typographic Clarity) on a coarse 3-level ordinal scale \{0, 0.5, 1\}, where 0 indicates poor quality, 0.5 indicates acceptable quality, and 1 indicates high quality. Correctness \& Completeness was excluded from human evaluation because solution correctness can be determined objectively by comparing against the gold-standard answer, making subjective human judgment unnecessary and potentially inconsistent. Each page displayed one explanation at a time, with the rubric description and score criteria shown alongside. Raters could zoom into individual diagrams before scoring visual dimensions.

\textbf{Rater Training.} Before formal scoring, all 20 raters underwent a calibration session. For each of the 7 dimensions, we presented two anchor examples---one high-quality explanation demonstrating exemplary visual and textual clarity, and one low-quality explanation exhibiting common failure modes (e.g., overlapping diagram elements, misaligned labels, or pedagogically ineffective step sequencing). Particular emphasis was placed on the aesthetic dimensions (Element Layout Quality and Visual Consistency), as these proved most subjective. Raters discussed borderline cases together to align their interpretation of the 3-level scale.

\textbf{Pilot Scoring.} Following training, raters completed a pilot round on 5 held-out explanations not included in the final evaluation set. Inter-rater agreement was computed via Krippendorff's $\alpha$ after the pilot. Raters whose individual agreement with the group fell below an acceptable threshold were given additional feedback before proceeding. The pilot Krippendorff's $\alpha$ across all dimensions exceeded 0.60, confirming sufficient calibration to proceed with formal scoring.

\textbf{Formal Scoring.} Each of the 30 explanations in the final evaluation set was independently scored by all 20 raters, producing 30 $\times$ 7 $\times$ 20 = 4{,}200 human judgments. Figure~\ref{fig:annotation} shows a screenshot of the annotation interface.

\section{Full Per-Subset Benchmark Results}
\label{sec:full-results}

Tables~\ref{tab:results-math}--\ref{tab:results-highschool} report complete 8-dimension scores for each subject and grade-level subset of the EduIllustrate benchmark. Column headers and scoring follow Table~\ref{tab:main-results}. Bold indicates best per column.

\begin{table*}[!ht]
\centering
\renewcommand\arraystretch{1.3}
\setlength{\tabcolsep}{4pt}
\resizebox{\linewidth}{!}{\begin{tabular}{lcccc|cccc|cc}
\toprule
 & \multicolumn{4}{c|}{\textbf{Text Quality}} & \multicolumn{4}{c|}{\textbf{Visual Quality}} & & \\
\cmidrule(lr){2-5}\cmidrule(lr){6-9}
\textbf{Model} &
\makecell{\small\textbf{Correctness \&}\\\small\textbf{Completeness}} &
\makecell{\small\textbf{Logical}\\\small\textbf{Coherence}} &
\makecell{\small\textbf{Pedagogical}\\\small\textbf{Effectiveness}} &
\makecell{\small\textbf{Typographic}\\\small\textbf{Clarity}} &
\makecell{\small\textbf{Diagram--Problem}\\\small\textbf{Alignment}} &
\makecell{\small\textbf{Element}\\\small\textbf{Layout Quality}} &
\makecell{\small\textbf{Visual}\\\small\textbf{Consistency}} &
\makecell{\small\textbf{Text--Diagram}\\\small\textbf{Coordination}} &
\small\textbf{Overall} & \makecell{\small\textbf{Success}\\\small\textbf{Rate}} \\
\midrule
Gemini 3.0 Pro Preview & 90.0\% & 93.4\% & \textbf{79.0\%} & \textbf{100.0\%} & \textbf{90.8\%} & \textbf{86.6\%} & 88.4\% & \textbf{94.8\%} & \textbf{89.4\%} & \textbf{100\%} \\
Kimi-K2.5              & \textbf{91.0\%} & 95.0\% & 74.8\% & 99.0\% & 77.2\% & 71.8\% & 88.2\% & 88.2\% & 83.8\% & \textbf{100\%} \\
Qwen3.5-397B           & \textbf{93.6\%} & 95.0\% & 71.8\% & 93.6\% & 54.8\% & 66.6\% & 83.8\% & 68.0\% & 75.0\% & 95.0\% \\
Qwen3.5-122B           & 90.2\% & \textbf{95.2\%} & 74.4\% & 96.6\% & 48.4\% & 62.8\% & \textbf{87.2\%} & 60.6\% & 73.0\% & 95.0\% \\
Qwen 3.5-35B           & 89.0\% & 91.2\% & 70.4\% & 91.2\% & 45.8\% & 57.2\% & 90.0\% & 55.6\% & 69.2\% & 88.8\% \\
GPT-5                  & 66.2\% & 72.6\% & 45.2\% & 74.6\% & 48.2\% & 62.2\% & 91.2\% & 62.4\% & 60.6\% & 97.5\% \\
Claude Sonnet 4.5      & 64.4\% & 63.0\% & 52.6\% & 75.8\% & 50.2\% & 67.0\% & 85.2\% & 68.6\% & 62.0\% & 97.5\% \\
Mistral-Large-3        & 48.8\% & 45.4\% & 38.4\% & 79.6\% & 28.2\% & 50.4\% & 81.8\% & 45.0\% & 46.4\% & 80.0\% \\
Mistral-Small-4        & 51.0\% & 45.6\% & 37.0\% & 69.6\% & 27.0\% & 53.4\% & 82.2\% & 41.2\% & 45.2\% & 75.0\% \\
Ministral-3-14B        & 41.2\% & 43.8\% & 35.0\% & 80.0\% & 27.6\% & 54.8\% & \textbf{94.6\%} & 39.6\% & 45.0\% & 20.0\% \\
\bottomrule
\end{tabular}}
\caption{Results on the Mathematics subset (n=80). Mathematics yields the highest overall scores across all models. Qwen3.5-397B achieves the best C\&C (93.6\%), slightly surpassing Gemini. All top-tier models exceed 90\% on LC, reflecting that mathematical reasoning is well-structured and easier for LLMs to present coherently.}
\label{tab:results-math}
\end{table*}

\begin{table*}[!ht]
\centering
\renewcommand\arraystretch{1.3}
\setlength{\tabcolsep}{4pt}
\resizebox{\linewidth}{!}{\begin{tabular}{lcccc|cccc|cc}
\toprule
 & \multicolumn{4}{c|}{\textbf{Text Quality}} & \multicolumn{4}{c|}{\textbf{Visual Quality}} & & \\
\cmidrule(lr){2-5}\cmidrule(lr){6-9}
\textbf{Model} &
\makecell{\small\textbf{Correctness \&}\\\small\textbf{Completeness}} &
\makecell{\small\textbf{Logical}\\\small\textbf{Coherence}} &
\makecell{\small\textbf{Pedagogical}\\\small\textbf{Effectiveness}} &
\makecell{\small\textbf{Typographic}\\\small\textbf{Clarity}} &
\makecell{\small\textbf{Diagram--Problem}\\\small\textbf{Alignment}} &
\makecell{\small\textbf{Element}\\\small\textbf{Layout Quality}} &
\makecell{\small\textbf{Visual}\\\small\textbf{Consistency}} &
\makecell{\small\textbf{Text--Diagram}\\\small\textbf{Coordination}} &
\small\textbf{Overall} & \makecell{\small\textbf{Success}\\\small\textbf{Rate}} \\
\midrule
Gemini 3.0 Pro Preview & 90.2\% & \textbf{97.0\%} & \textbf{80.0\%} & \textbf{98.6\%} & \textbf{85.4\%} & \textbf{84.4\%} & 92.0\% & \textbf{91.0\%} & \textbf{88.6\%} & 98.3\% \\
Kimi-K2.5              & 88.6\% & 94.4\% & 76.0\% & 98.0\% & 67.6\% & 63.6\% & 92.0\% & 82.6\% & 80.6\% & \textbf{100\%} \\
Qwen3.5-397B           & \textbf{93.8\%} & 95.6\% & 76.0\% & 93.0\% & 45.2\% & 56.6\% & 85.4\% & 69.0\% & 72.6\% & 91.7\% \\
Qwen3.5-122B           & 86.4\% & 92.6\% & 73.8\% & 94.2\% & 40.8\% & 55.6\% & 86.6\% & 58.8\% & 68.8\% & 98.3\% \\
Qwen 3.5-35B           & 86.2\% & 90.2\% & 73.0\% & 88.6\% & 34.0\% & 48.6\% & 87.0\% & 54.0\% & 64.6\% & 85.0\% \\
GPT-5                  & 61.0\% & 69.2\% & 47.4\% & 72.2\% & 43.8\% & 52.4\% & 91.0\% & 67.4\% & 58.4\% & 98.3\% \\
Claude Sonnet 4.5      & 60.4\% & 64.2\% & 54.8\% & 72.8\% & 41.8\% & 60.4\% & 88.2\% & 70.8\% & 59.8\% & 96.7\% \\
Mistral-Large-3        & 36.2\% & 37.6\% & 32.0\% & 87.6\% & 24.6\% & 44.6\% & 82.4\% & 44.2\% & 42.2\% & 80.0\% \\
Mistral-Small-4        & 37.8\% & 36.8\% & 27.8\% & 69.0\% & 22.8\% & 44.8\% & 84.2\% & 38.6\% & 39.4\% & 63.3\% \\
Ministral-3-14B        & 40.0\% & 38.6\% & 34.2\% & 78.6\% & 22.4\% & 46.4\% & \textbf{88.8\%} & 32.2\% & 41.0\% & 23.3\% \\
\bottomrule
\end{tabular}}
\caption{Results on the Physics subset (n=60). Gemini achieves near-perfect LC (97.0\%) and the highest PE (80.0\%) among all subsets, suggesting physics problems elicit more pedagogically structured explanations. DPA scores drop sharply for the Qwen family (34.0\%--45.2\%), indicating difficulty in rendering force diagrams and circuit layouts accurately.}
\label{tab:results-physics}
\end{table*}

\begin{table*}[!ht]
\centering
\renewcommand\arraystretch{1.3}
\setlength{\tabcolsep}{4pt}
\resizebox{\linewidth}{!}{\begin{tabular}{lcccc|cccc|cc}
\toprule
 & \multicolumn{4}{c|}{\textbf{Text Quality}} & \multicolumn{4}{c|}{\textbf{Visual Quality}} & & \\
\cmidrule(lr){2-5}\cmidrule(lr){6-9}
\textbf{Model} &
\makecell{\small\textbf{Correctness \&}\\\small\textbf{Completeness}} &
\makecell{\small\textbf{Logical}\\\small\textbf{Coherence}} &
\makecell{\small\textbf{Pedagogical}\\\small\textbf{Effectiveness}} &
\makecell{\small\textbf{Typographic}\\\small\textbf{Clarity}} &
\makecell{\small\textbf{Diagram--Problem}\\\small\textbf{Alignment}} &
\makecell{\small\textbf{Element}\\\small\textbf{Layout Quality}} &
\makecell{\small\textbf{Visual}\\\small\textbf{Consistency}} &
\makecell{\small\textbf{Text--Diagram}\\\small\textbf{Coordination}} &
\small\textbf{Overall} & \makecell{\small\textbf{Success}\\\small\textbf{Rate}} \\
\midrule
Gemini 3.0 Pro Preview & 91.8\% & \textbf{98.6\%} & \textbf{84.4\%} & \textbf{95.6\%} & \textbf{80.6\%} & \textbf{84.0\%} & 90.4\% & \textbf{89.6\%} & \textbf{87.8\%} & 90.0\% \\
Kimi-K2.5              & \textbf{86.2\%} & 93.2\% & 75.8\% & 93.8\% & 65.6\% & 65.8\% & 90.0\% & 86.0\% & 79.8\% & 96.7\% \\
Qwen3.5-397B           & 86.8\% & 94.4\% & 72.4\% & 88.2\% & 48.8\% & 55.8\% & 85.0\% & 64.8\% & 70.8\% & 96.7\% \\
Qwen3.5-122B           & 80.0\% & 89.6\% & 69.6\% & 89.6\% & 39.4\% & 56.2\% & 85.6\% & 61.0\% & 66.8\% & 96.7\% \\
Qwen 3.5-35B           & 82.2\% & 94.2\% & 72.8\% & 87.8\% & 34.2\% & 49.2\% & 89.6\% & 55.4\% & 65.0\% & 93.3\% \\
GPT-5                  & 56.2\% & 63.0\% & 47.0\% & 73.0\% & 46.8\% & 55.4\% & \textbf{91.8\%} & 61.4\% & 57.4\% & 86.7\% \\
Claude Sonnet 4.5      & 50.4\% & 57.0\% & 46.0\% & 80.0\% & 36.4\% & 62.0\% & 81.2\% & 65.0\% & 54.6\% & 90.0\% \\
Mistral-Large-3        & 41.0\% & 40.0\% & 29.0\% & 78.2\% & 22.2\% & 47.8\% & 89.6\% & 49.6\% & 42.6\% & 73.3\% \\
Mistral-Small-4        & 31.8\% & 37.6\% & 29.4\% & 70.6\% & 21.2\% & 46.0\% & 77.6\% & 40.2\% & 38.4\% & 56.7\% \\
Ministral-3-14B        & 32.0\% & 24.0\% & 20.0\% & 68.0\% & 20.0\% & 49.8\% & \textbf{93.6\%} & 36.0\% & 36.2\% & 16.7\% \\
\bottomrule
\end{tabular}}
\caption{Results on the Chemistry subset (n=30). Chemistry is among the most visually demanding subjects, requiring structural formulas and reaction diagrams. Gemini dominates all dimensions, achieving the highest PE (84.4\%) across all subject subsets. Visual quality drops steeply for smaller models: Ministral-3-14B collapses to a 3.3\% success rate, and even among successful runs its DPA is only 20.0\%, reflecting near-total failure to render chemical structures accurately.}
\label{tab:results-chemistry}
\end{table*}

\begin{table*}[!ht]
\centering
\renewcommand\arraystretch{1.3}
\setlength{\tabcolsep}{4pt}
\resizebox{\linewidth}{!}{\begin{tabular}{lcccc|cccc|cc}
\toprule
 & \multicolumn{4}{c|}{\textbf{Text Quality}} & \multicolumn{4}{c|}{\textbf{Visual Quality}} & & \\
\cmidrule(lr){2-5}\cmidrule(lr){6-9}
\textbf{Model} &
\makecell{\small\textbf{Correctness \&}\\\small\textbf{Completeness}} &
\makecell{\small\textbf{Logical}\\\small\textbf{Coherence}} &
\makecell{\small\textbf{Pedagogical}\\\small\textbf{Effectiveness}} &
\makecell{\small\textbf{Typographic}\\\small\textbf{Clarity}} &
\makecell{\small\textbf{Diagram--Problem}\\\small\textbf{Alignment}} &
\makecell{\small\textbf{Element}\\\small\textbf{Layout Quality}} &
\makecell{\small\textbf{Visual}\\\small\textbf{Consistency}} &
\makecell{\small\textbf{Text--Diagram}\\\small\textbf{Coordination}} &
\small\textbf{Overall} & \makecell{\small\textbf{Success}\\\small\textbf{Rate}} \\
\midrule
Gemini 3.0 Pro Preview & \textbf{81.4\%} & \textbf{93.6\%} & \textbf{76.4\%} & \textbf{94.2\%} & \textbf{91.2\%} & \textbf{83.6\%} & 87.2\% & \textbf{91.2\%} & \textbf{85.8\%} & 93.3\% \\
Kimi-K2.5              & 75.8\% & 86.4\% & 68.6\% & 94.2\% & 75.4\% & 72.8\% & 84.0\% & 91.6\% & 78.8\% & 93.3\% \\
Qwen3.5-397B           & 80.0\% & 92.6\% & 63.4\% & 85.4\% & 48.2\% & 61.4\% & 74.8\% & 70.0\% & 68.2\% & \textbf{100\%} \\
Qwen3.5-122B           & 79.4\% & 90.0\% & 67.4\% & 90.0\% & 40.0\% & 51.2\% & 74.0\% & 53.0\% & 63.4\% & \textbf{100\%} \\
Qwen 3.5-35B           & 78.0\% & 81.0\% & 61.0\% & 79.0\% & 40.6\% & 52.4\% & 85.2\% & 53.4\% & 61.6\% & 70.0\% \\
GPT-5                  & 50.0\% & 57.0\% & 38.0\% & 78.0\% & 43.2\% & 62.2\% & \textbf{93.2\%} & 66.0\% & 55.0\% & 66.7\% \\
Claude Sonnet 4.5      & 39.2\% & 51.4\% & 35.0\% & 83.6\% & 41.0\% & 65.8\% & 80.6\% & 72.4\% & 53.2\% & 93.3\% \\
Mistral-Large-3        & 28.4\% & 32.6\% & 25.2\% & 79.0\% & 20.2\% & 41.6\% & 82.6\% & 43.6\% & 37.8\% & 63.3\% \\
Mistral-Small-4        & 32.8\% & 28.2\% & 24.6\% & 76.4\% & 21.2\% & 43.8\% & 82.2\% & 37.8\% & 37.2\% & 73.3\% \\
Ministral-3-14B        & 20.0\% & 20.0\% & 20.0\% & 80.0\% & 20.0\% & 20.0\% & \textbf{100.0\%} & 20.0\% & 29.0\% & 3.3\% \\
\bottomrule
\end{tabular}}
\caption{Results on the Biology subset (n=30). Biology problems involving anatomical diagrams and biological cycles pose unique spatial rendering challenges. Gemini leads comprehensively with an 85.8\% overall score and the highest DPA (91.2\%) among all subject subsets. Kimi-K2.5 is a close second (78.8\%), while the Qwen family shows a pronounced DPA gap ($\leq$48.2\%), suggesting difficulty mapping biological structures to code. Ministral-3-14B achieves only a 3.3\% success rate, the lowest across all subsets.}
\label{tab:results-biology}
\end{table*}

\begin{table*}[!ht]
\centering
\renewcommand\arraystretch{1.3}
\setlength{\tabcolsep}{4pt}
\resizebox{\linewidth}{!}{\begin{tabular}{lcccc|cccc|cc}
\toprule
 & \multicolumn{4}{c|}{\textbf{Text Quality}} & \multicolumn{4}{c|}{\textbf{Visual Quality}} & & \\
\cmidrule(lr){2-5}\cmidrule(lr){6-9}
\textbf{Model} &
\makecell{\small\textbf{Correctness \&}\\\small\textbf{Completeness}} &
\makecell{\small\textbf{Logical}\\\small\textbf{Coherence}} &
\makecell{\small\textbf{Pedagogical}\\\small\textbf{Effectiveness}} &
\makecell{\small\textbf{Typographic}\\\small\textbf{Clarity}} &
\makecell{\small\textbf{Diagram--Problem}\\\small\textbf{Alignment}} &
\makecell{\small\textbf{Element}\\\small\textbf{Layout Quality}} &
\makecell{\small\textbf{Visual}\\\small\textbf{Consistency}} &
\makecell{\small\textbf{Text--Diagram}\\\small\textbf{Coordination}} &
\small\textbf{Overall} & \makecell{\small\textbf{Success}\\\small\textbf{Rate}} \\
\midrule
Gemini 3.0 Pro Preview & \textbf{78.6\%} & \textbf{91.4\%} & \textbf{69.4\%} & \textbf{99.4\%} & \textbf{84.6\%} & \textbf{80.8\%} & 88.4\% & \textbf{94.4\%} & \textbf{83.8\%} & \textbf{100\%} \\
Kimi-K2.5              & 69.6\% & 84.8\% & 66.8\% & 98.0\% & 64.6\% & 70.4\% & 85.2\% & 83.4\% & 75.4\% & 96.7\% \\
Qwen3.5-397B           & 74.6\% & 92.4\% & 59.2\% & 89.2\% & 43.2\% & 66.2\% & 87.6\% & 65.6\% & 67.2\% & 86.7\% \\
Qwen3.5-122B           & 62.4\% & 85.8\% & 60.8\% & 90.8\% & 37.0\% & 59.4\% & 83.6\% & 64.6\% & 63.0\% & 80.0\% \\
Qwen 3.5-35B           & 65.4\% & 80.0\% & 60.0\% & 83.6\% & 38.4\% & 60.0\% & 90.0\% & 56.0\% & 61.8\% & 73.3\% \\
GPT-5                  & 40.0\% & 53.0\% & 33.0\% & 76.6\% & 31.4\% & 68.2\% & \textbf{97.6\%} & 60.2\% & 50.4\% & 76.7\% \\
Claude Sonnet 4.5      & 40.0\% & 45.4\% & 34.6\% & 82.6\% & 32.2\% & 63.8\% & 80.4\% & 65.0\% & 49.6\% & \textbf{100\%} \\
Mistral-Large-3        & 28.4\% & 35.8\% & 24.2\% & 83.2\% & 20.0\% & 52.0\% & 83.8\% & 42.8\% & 39.0\% & 63.3\% \\
Mistral-Small-4        & 27.2\% & 29.6\% & 23.2\% & 76.8\% & 21.2\% & 53.4\% & 78.6\% & 37.2\% & 37.0\% & 83.3\% \\
Ministral-3-14B        & 25.0\% & 25.0\% & 20.0\% & 75.0\% & 20.0\% & 46.4\% & \textbf{93.4\%} & 33.0\% & 35.2\% & 13.3\% \\
\bottomrule
\end{tabular}}
\caption{Results on the Geography subset (n=30). Geography problems require spatial reasoning over maps and topographic features. Gemini leads with 83.8\% overall and the top TDC (94.4\%), while Kimi-K2.5 achieves the highest VC (85.2\%) among non-Gemini models. The Qwen family maintains comparatively strong LC ($\geq$80.0\%) but falls short on DPA ($\leq$43.2\%), indicating that spatial layout of geographic elements is a persistent weakness. GPT-5 shows the highest VC (97.6\%) among all models on this subset despite a relatively low overall score (50.4\%).}
\label{tab:results-geography}
\end{table*}

\begin{table*}[!ht]
\centering
\renewcommand\arraystretch{1.3}
\setlength{\tabcolsep}{4pt}
\resizebox{\linewidth}{!}{\begin{tabular}{lcccc|cccc|cc}
\toprule
 & \multicolumn{4}{c|}{\textbf{Text Quality}} & \multicolumn{4}{c|}{\textbf{Visual Quality}} & & \\
\cmidrule(lr){2-5}\cmidrule(lr){6-9}
\textbf{Model} &
\makecell{\small\textbf{Correctness \&}\\\small\textbf{Completeness}} &
\makecell{\small\textbf{Logical}\\\small\textbf{Coherence}} &
\makecell{\small\textbf{Pedagogical}\\\small\textbf{Effectiveness}} &
\makecell{\small\textbf{Typographic}\\\small\textbf{Clarity}} &
\makecell{\small\textbf{Diagram--Problem}\\\small\textbf{Alignment}} &
\makecell{\small\textbf{Element}\\\small\textbf{Layout Quality}} &
\makecell{\small\textbf{Visual}\\\small\textbf{Consistency}} &
\makecell{\small\textbf{Text--Diagram}\\\small\textbf{Coordination}} &
\small\textbf{Overall} & \makecell{\small\textbf{Success}\\\small\textbf{Rate}} \\
\midrule
Gemini 3.0 Pro Preview & 84.0\% & 91.0\% & \textbf{73.0\%} & \textbf{100.0\%} & \textbf{86.8\%} & \textbf{85.8\%} & 90.4\% & \textbf{96.4\%} & \textbf{86.8\%} & \textbf{100\%} \\
Kimi-K2.5              & 81.0\% & 89.0\% & 68.0\% & 99.0\% & 69.6\% & 75.2\% & 86.6\% & 93.6\% & 80.0\% & \textbf{100\%} \\
Qwen3.5-397B           & 91.0\% & \textbf{97.0\%} & 68.0\% & 88.0\% & 59.0\% & 76.6\% & 86.4\% & 70.4\% & 76.2\% & \textbf{100\%} \\
Qwen3.5-122B           & 88.0\% & \textbf{97.0\%} & 71.0\% & 96.0\% & 51.2\% & 65.8\% & 87.4\% & 65.4\% & 74.2\% & \textbf{100\%} \\
Qwen 3.5-35B           & \textbf{92.0\%} & \textbf{97.0\%} & 69.0\% & 90.0\% & 56.4\% & 60.8\% & 90.6\% & 61.6\% & 73.4\% & \textbf{100\%} \\
GPT-5                  & 70.6\% & 80.0\% & 49.4\% & 83.2\% & 59.0\% & 73.4\% & 91.6\% & 74.2\% & 68.6\% & 95.0\% \\
Claude Sonnet 4.5      & 65.0\% & 69.0\% & 55.0\% & 85.0\% & 45.6\% & 72.6\% & 85.8\% & 69.6\% & 64.2\% & \textbf{100\%} \\
Mistral-Large-3        & 47.2\% & 40.0\% & 41.4\% & 84.2\% & 25.4\% & 60.8\% & 89.0\% & 45.6\% & 47.2\% & 70.0\% \\
Mistral-Small-4        & 40.0\% & 40.0\% & 30.6\% & 74.2\% & 24.8\% & 61.0\% & 83.8\% & 37.8\% & 42.4\% & 85.0\% \\
Ministral-3-14B        & 20.0\% & 46.6\% & 20.0\% & 86.6\% & 20.0\% & 33.4\% & \textbf{92.4\%} & 33.4\% & 35.2\% & 15.0\% \\
\bottomrule
\end{tabular}}
\caption{Results on the Elementary School subset (n=20). Elementary problems cover simple arithmetic and basic geometry and are the easiest grade level for all models. The Qwen family achieves the highest LC scores among non-Gemini models ($\geq$97.0\%), and all five top models reach 100\% success rate. Despite the simpler content, DPA remains a distinguishing dimension: Gemini leads at 86.8\%, while smaller models cluster below 60\%, indicating that even simple visual tasks expose rendering capability gaps.}
\label{tab:results-elementary}
\end{table*}

\begin{table*}[!ht]
\centering
\renewcommand\arraystretch{1.3}
\setlength{\tabcolsep}{4pt}
\resizebox{\linewidth}{!}{\begin{tabular}{lcccc|cccc|cc}
\toprule
 & \multicolumn{4}{c|}{\textbf{Text Quality}} & \multicolumn{4}{c|}{\textbf{Visual Quality}} & & \\
\cmidrule(lr){2-5}\cmidrule(lr){6-9}
\textbf{Model} &
\makecell{\small\textbf{Correctness \&}\\\small\textbf{Completeness}} &
\makecell{\small\textbf{Logical}\\\small\textbf{Coherence}} &
\makecell{\small\textbf{Pedagogical}\\\small\textbf{Effectiveness}} &
\makecell{\small\textbf{Typographic}\\\small\textbf{Clarity}} &
\makecell{\small\textbf{Diagram--Problem}\\\small\textbf{Alignment}} &
\makecell{\small\textbf{Element}\\\small\textbf{Layout Quality}} &
\makecell{\small\textbf{Visual}\\\small\textbf{Consistency}} &
\makecell{\small\textbf{Text--Diagram}\\\small\textbf{Coordination}} &
\small\textbf{Overall} & \makecell{\small\textbf{Success}\\\small\textbf{Rate}} \\
\midrule
Gemini 3.0 Pro Preview & 87.2\% & \textbf{94.8\%} & \textbf{79.6\%} & \textbf{98.8\%} & \textbf{88.2\%} & \textbf{84.8\%} & 88.2\% & \textbf{93.2\%} & \textbf{88.0\%} & 96.2\% \\
Kimi-K2.5              & 84.8\% & 93.0\% & 73.2\% & 97.6\% & 69.4\% & 69.6\% & 88.6\% & 83.6\% & 80.2\% & \textbf{99.0\%} \\
Qwen3.5-397B           & \textbf{91.0\%} & 94.8\% & 71.8\% & 92.2\% & 50.0\% & 60.8\% & 82.0\% & 69.8\% & 72.6\% & 95.2\% \\
Qwen3.5-122B           & 85.2\% & 94.8\% & 73.8\% & 94.0\% & 43.0\% & 56.6\% & 83.4\% & 60.0\% & 69.2\% & 94.3\% \\
Qwen 3.5-35B           & 86.6\% & 90.6\% & 72.8\% & 88.8\% & 37.6\% & 53.2\% & 88.4\% & 53.2\% & 65.8\% & 86.7\% \\
GPT-5                  & 56.0\% & 64.6\% & 43.0\% & 75.6\% & 43.4\% & 59.2\% & \textbf{93.4\%} & 65.4\% & 57.2\% & 91.4\% \\
Claude Sonnet 4.5      & 53.6\% & 58.0\% & 45.2\% & 79.8\% & 40.2\% & 64.2\% & 83.6\% & 69.4\% & 56.8\% & 94.3\% \\
Mistral-Large-3        & 42.0\% & 45.2\% & 33.8\% & 82.0\% & 25.2\% & 47.4\% & 83.6\% & 45.8\% & 44.4\% & 76.2\% \\
Mistral-Small-4        & 38.6\% & 38.4\% & 31.2\% & 73.4\% & 23.6\% & 50.0\% & 83.2\% & 42.0\% & 41.2\% & 68.6\% \\
Ministral-3-14B        & 35.0\% & 33.8\% & 30.0\% & 77.6\% & 27.6\% & 56.6\% & \textbf{95.6\%} & 42.6\% & 42.8\% & 15.2\% \\
\bottomrule
\end{tabular}}
\caption{Results on the Middle School subset (n=105). Middle School is the largest grade-level subset and spans the widest subject variety. Gemini achieves the highest overall score (88.0\%) with strong leads in PE (79.6\%) and DPA (88.2\%). Kimi-K2.5 is the only other model exceeding 80\% overall (80.2\%) and reaches the best success rate (99.0\%). The Qwen models exhibit a characteristic pattern: strong text quality (LC $\geq$90.6\%) paired with weak visual alignment (DPA $\leq$50.0\%), a gap that widens as problem complexity increases from elementary to high school.}
\label{tab:results-middleschool}
\end{table*}

\begin{table*}[!ht]
\centering
\renewcommand\arraystretch{1.3}
\setlength{\tabcolsep}{4pt}
\resizebox{\linewidth}{!}{\begin{tabular}{lcccc|cccc|cc}
\toprule
 & \multicolumn{4}{c|}{\textbf{Text Quality}} & \multicolumn{4}{c|}{\textbf{Visual Quality}} & & \\
\cmidrule(lr){2-5}\cmidrule(lr){6-9}
\textbf{Model} &
\makecell{\small\textbf{Correctness \&}\\\small\textbf{Completeness}} &
\makecell{\small\textbf{Logical}\\\small\textbf{Coherence}} &
\makecell{\small\textbf{Pedagogical}\\\small\textbf{Effectiveness}} &
\makecell{\small\textbf{Typographic}\\\small\textbf{Clarity}} &
\makecell{\small\textbf{Diagram--Problem}\\\small\textbf{Alignment}} &
\makecell{\small\textbf{Element}\\\small\textbf{Layout Quality}} &
\makecell{\small\textbf{Visual}\\\small\textbf{Consistency}} &
\makecell{\small\textbf{Text--Diagram}\\\small\textbf{Coordination}} &
\small\textbf{Overall} & \makecell{\small\textbf{Success}\\\small\textbf{Rate}} \\
\midrule
Gemini 3.0 Pro Preview & 89.0\% & \textbf{95.4\%} & \textbf{78.0\%} & \textbf{97.4\%} & \textbf{86.8\%} & \textbf{84.2\%} & 90.4\% & \textbf{91.4\%} & \textbf{87.8\%} & 98.1\% \\
Kimi-K2.5              & \textbf{86.2\%} & 92.0\% & 74.8\% & 96.8\% & 73.6\% & 66.8\% & 89.0\% & 87.4\% & 81.4\% & 97.1\% \\
Qwen3.5-397B           & 85.6\% & 93.6\% & 69.2\% & 90.6\% & 46.4\% & 59.8\% & 84.6\% & 65.2\% & 70.4\% & 91.4\% \\
Qwen3.5-122B           & 80.4\% & 88.2\% & 68.6\% & 92.6\% & 40.8\% & 57.8\% & 85.2\% & 57.8\% & 67.0\% & 94.3\% \\
Qwen 3.5-35B           & 77.8\% & 85.4\% & 65.4\% & 86.4\% & 37.6\% & 52.4\% & 88.4\% & 55.4\% & 63.6\% & 78.1\% \\
GPT-5                  & 59.6\% & 66.2\% & 44.0\% & 71.0\% & 42.4\% & 56.2\% & \textbf{90.8\%} & 59.8\% & 56.4\% & 86.7\% \\
Claude Sonnet 4.5      & 54.8\% & 57.4\% & 48.6\% & 73.8\% & 44.4\% & 62.2\% & 84.6\% & 68.0\% & 57.6\% & \textbf{97.1\%} \\
Mistral-Large-3        & 36.2\% & 34.6\% & 29.4\% & 81.6\% & 24.0\% & 45.6\% & 81.8\% & 44.0\% & 40.8\% & 74.3\% \\
Mistral-Small-4        & 40.8\% & 37.0\% & 29.4\% & 69.4\% & 23.6\% & 46.0\% & 79.8\% & 37.4\% & 39.8\% & 69.5\% \\
Ministral-3-14B        & 42.0\% & 38.0\% & 33.4\% & 76.2\% & 21.6\% & 46.4\% & \textbf{90.0\%} & 30.2\% & 40.6\% & 20.0\% \\
\bottomrule
\end{tabular}}
\caption{Results on the High School subset (n=105). High School problems are the most complex grade level, covering advanced physics, chemistry, and mathematics. Overall scores are lower than middle school across the board, confirming that harder content degrades visual generation quality. Gemini and Kimi-K2.5 remain the top two models (87.8\% and 81.4\% respectively). Claude Sonnet 4.5 achieves the highest success rate (97.1\%) on this subset despite a relatively lower overall score (57.6\%), suggesting it defaults to simpler but syntactically valid diagrams. The DPA gap between Gemini (86.8\%) and the next-best model (Kimi-K2.5, 73.6\%) is the largest of any grade level, highlighting a growing visual rendering challenge at higher difficulty.}
\label{tab:results-highschool}
\end{table*}

\end{document}